\documentclass{article}
\usepackage{arxiv}
\usepackage{url}
\usepackage{multirow}
\usepackage{textcomp}
\usepackage{algorithm}
\usepackage[utf8]{inputenc}
\usepackage[T1]{fontenc}	 
\usepackage[noend]{algpseudocode}
\usepackage{graphicx}
\usepackage{epstopdf}
\usepackage{subfloat}
\usepackage{subfig}
\usepackage{amsmath}
\usepackage{amssymb}
\usepackage{multicol}
\usepackage{authblk}
\usepackage{graphicx}
\usepackage[utf8]{inputenc} 
\usepackage[T1]{fontenc}    
\usepackage{url}            
\usepackage{booktabs}       
\usepackage{amsfonts}       
\usepackage{nicefrac}       
\usepackage{microtype}      
\usepackage{lipsum}
\usepackage{rotating}
\title{A Survey on Ethical Hacking: Issues and Challenges}

\author{
 Jean-Paul A. Yaacoub, Hassan N. Noura,  Ola Salman, Ali Chehab \\
   American University of Beirut,\\
   Electrical and Computer Engineering Department,\\ Beirut 1107 2020, Lebanon\\
  
}
\begin{document}

\maketitle
\begin{abstract}
Security attacks are growing in an exponential manner and their impact on existing systems is seriously high and can lead to dangerous consequences. However, in order to reduce the effect of these attacks, penetration tests are highly required, and can be considered as a suitable solution for this task. Therefore, the main focus of this paper is to explain the technical and non-technical steps of penetration tests. The objective of penetration tests is to make existing systems and their corresponding data more secure, efficient and resilient. In other terms, pen testing is a simulated attack with the goal of identifying any exploitable vulnerability or/and a security gap. In fact, any identified exploitable vulnerability will be used to conduct attacks on systems, devices, or personnel. This growing problem should be solved and mitigated to reach better resistance against these attacks. Moreover, the advantages and limitations of penetration tests are also listed. The main issue of penetration tests that it is efficient to detect known vulnerabilities. Therefore, in order to resist unknown vulnerabilities, a new kind of modern penetration tests is required, in addition to reinforcing the use of shadows honeypots. This can also be done by reinforcing the anomaly detection of intrusion detection/prevention system. In fact, security is increased by designing an efficient cooperation between the different security elements and penetration tests. 

\end{abstract}



\section{Introduction}

Due to the current involvement of the digital world in our daily lives and the huge reliance on it, the protection of a given system can be less efficient as a result of poor or absence of the right security measures. As a result, malicious hackers can exploit these vulnerabilities and security gaps. In fact, most of the currently available websites, networks, and applications are poorly and hastily configured. This lack of security is mainly caused by poor planning and/or poor coding configurations. This means that no consideration is taken about the consequences of any attack. Thus, making them prone to various types of malicious code injection attacks, along with networks, applications, web services, and infrastructure attacks. Therefore, this paper discusses how these attack are conducted, along how they can affect one, many or all of the security goals in a given system. Furthermore, the lack of training and awareness among employees and IT staff is also problematic and seriously challenging. In fact, there is a remarkable and significant skill gap in the ethical hacking domain. Therefore, the need to continuously and constantly train employees, along with the need for more ethical hackers is a must. Additionally, this paper also highlights the need for new security measures. This has led pen testing to become a new emerging trend, which is achieved by evaluating both levels of security and immunity against already known attacks and threats. This is done through vulnerability assessment, foot-printing, risk evaluation, and pen testing.  


The main aim in this paper is to highlight the importance of performing and conducting penetration testing~\cite{bishop2007penetration}. This helps in evaluating the levels of security and immunity of a given organisation against already known threats, risks and attacks. This can be achieved through assessment and mitigation. In fact, it is recommended to identify the main difference between hackers and ethical hackers. The importance of ethical hackers~\cite{murad2008certified} is also highlighted in the cyber-domain. This helps raising the levels of security and awareness among different organisations (especially small organisations and businesses). This avoids falling victims of a given cyber-attack. Therefore, the goal is to encourage the use of pen testing in order to detect any exploitable vulnerability and security gap and fix it before any hacking attempt(s). In fact, it is an ongoing back-and-forth search-and-destroy domination race between hackers and ethical hackers over defending and attacking a targeted organisation. 


This paper presents a new holistic analytical view point in terms of understanding better the ethical hacking and pen testing domains, respectively. In fact, this paper identifies the main attacks that can be mitigated through pen testing and presents them with their modern countermeasures, unlike other papers that solely focused on SQL/malicious code injection prevention~\cite{kumar2012survey,kindy2011survey,mitropoulos2017fatal}, buffer-overflow protection~\cite{piromsopa2011survey}, botnet detection/prevention~\cite{zhang2011survey}, Distributed Denial-of-Service (DDoS), flooding and sniffing prevention~\cite{zhang2011survey,zargar2013survey,thakur2013content}, insider detection~\cite{salem2008survey}, and Trojan/worm detection~\cite{li2008survey,tehranipoor2010survey}. Moreover, this paper further studies ethical hackers’ motivations and motives to supplement the work done in~\cite{Ethicalh92:online}, and further support the choice of needing human hackers presented in~\cite{oakley2019human}. Finally, this paper is not limited to a single IoT aspect~\cite{ding2019ethical}, but instead it focuses on the important need of relying on ethical hackers and pen testers in all IoT aspects.

Thus, the contributions of this paper can be summarized as follows:
\begin{itemize}
\item \textbf{Methodical Way:} of classifying and analysing hackers based on their motives, activities, gains and their targets.
\item \textbf{Analytical Study:} of how to assess vulnerabilities and specify risks.
\item \textbf{Analysis:} of most known attacks, their structure, and how to mitigate them.
\item \textbf{Detailed Understanding:} on how pen testing is linked to ethical hackers, and how ethical hackers conduct their pen testing, along presenting different approaches from different authors.
\item \textbf{Evaluation:} of how pen testing is used to evaluate the security and protection levels against different possible potential attacks.
 \item \textbf{Security \& Safety Procedures:} are presented in a suitable way that maintain a secure environment pre/post and during an incident.
\end{itemize}


This paper is organised into nine sections including the introduction, and is presented as follows:
Section~\ref{sec:a} includes identifying and classifying hackers including their motives, gains, cyber-activity, along the most frequently targeted systems. Section~\ref{sec:2} assesses both technical and operational vulnerabilities, as well as systems' and personnel's vulnerabilities, along a study based on the risk specification. Section~\ref{sec:3} highlights the cyber-attack's structure, as well as the main cyber attack types such as web, application, network and infrastructure attacks. Section~\ref{sec:4} is divided into three main parts: the first part includes the concept of ethical hacking, along discussing its life-cycle, most used tools, as well as its main challenges and issues. The second part includes the main pen testing pros and cons, types, its phases, its most used tools, and its knowledge types, along its appliance. The last part discusses and analyses the already existing solutions. Section~\ref{sec:5} presents and discusses both security and safety features including their steps, as well as how incident responders react to an incident or/and event. Additionally, Section~\ref{sec:6} both preventing and protective security measures are discussed and analysed in accordance to maintaining the main security goals. Section~\ref{sec:7} presents the main learnt lessons and recommendations. Finally, section~\ref{sec:9} concludes the presented work.

\section{Background \& Overview}
\label{sec:a}

This section introduces the necessary  background and overview to further explain the processes of identification and classification of hackers. Moreover, it also highlights their motives, targets and gains through their performed cyber-activity in parallel with the adopted cyber-attack structure. Finally, a suitable framework will be presented.
\subsection{Identifying Hackers}
In the cyber-war of bytes and bits, it is important to know that a nation's white-hat hacker is another nation's black-hat hacker, and vice versa. This norm is presented and will be explained in this paper.
Before proceeding further, it is highly recommended to classify hackers. This classification depends on their motivation, knowledge, available resources, experience and skills. However, a brief comparison is presented in \tablename~\ref{tab:1} in order to highlight on the main differences between hackers and ethical hackers.
\begin{table*}[!ht]
\centering
\small
\caption{\textbf{Comparison Table Between Hackers \& Ethical Hackers}}
\label{tab:1}
\begin{tabular}{|l|l|l|}
\hline
\textbf{Differences} & \textbf{Hackers \& Crackers} & \textbf{Ethical Hackers} \\ \hline
 \textbf{Measures} &  {Offensive} &  {Defensive} \\ \hline
 \textbf{Attacks} &  {Harmful} &  {Simulated} \\ \hline
 \textbf{Vulnerabilities} &  {Exploited} &  {Identified} \\ \hline
 \textbf{Tools Used} &  {Same Tools} &  {Same Tools} \\ \hline
 \textbf{Purpose of Usage} &  {Malicious} &  {Non-malicious} \\ \hline
 \textbf{Organisations} &  {Attacking} &  {Protecting} \\ \hline
 \textbf{Security Evaluation} &  {Failed Attempts} &  {Pen Testing} \\ \hline
 \textbf{Security} &  {Breach} &  {Enhancement} \\ \hline
\end{tabular}
\end{table*}
 According to~\cite{hoffman2013hacker,levy1984hackers,shanmugapriya2013study}, hackers can be divided into seven main types.

 \begin{itemize}
\item \textbf{White-Hat:}
 are known as the good guys or "ethical hackers"~\cite{shanmugapriya2013study}. Their aim is to combat and fight against other hackers (mainly black-hat), in addition to raising the level of awareness among less tech savvy people through education and further constant training, workshops and conferences. Their main goal is to overcome cyber-attacks and reduce their impact. Hence, the extensive reliance and the use of penetration testing to help organisations identify their security gaps to help securing them. Certified Ethical Hackers (CEH)~\cite{gregg2006certified} are usually qualified in Offensive Security Certified Professional (OSCP), Council of Registered Security Testers (CREST),  Certified Infrastructure Tester (CIT) or CREST Certified Application Security Tester (CAST). In fact, some of the current white-hat hackers are former black hat hackers.

\item \textbf{Black Hat:}
are hackers with malicious intentions to cause harm as part of their pre-planned criminal act. The advanced version of black-hat hackers is known as ‘crackers’. Luckily, there's only a handful number of crackers. However, it is really hard to find them, as they are really experts in both, the knowledge and experience domains. Moreover, black hat hackers are capable of writing their own programs and even codes, after searching and finding vulnerabilities in a bank or organisation, and  exploiting them with the aim aim to either steal confidential information/money, cause financial damage or gain unauthorised access. 
\item \textbf{Grey Hat:}
 hackers can be classified as mercenaries, as they are capable of performing either offensive tasks (black-hat), or defensive tasks (white-hat) depending on whoever pays them the most~\cite{harris2004gray}. The grey hat category is among the most dangerous type of hackers, since they can quickly switch from black-hat hacking into white-hat hacking, and vice versa. 
\item \textbf{Green Hat:}
hackers are known as Neophyte or"n00bs", and in some cases they are known as "newbies". They are the youngest hackers having little or no knowledge and experience at all. Eventually, these hackers rely on the use of already written scripts, ask lots of questions and are typically very curious~\cite{levy1984hackers}.

\item \textbf{Red Hat:} 
hackers, are the ultra-white-hat hackers who use extreme and brute-force techniques and tactics to combat against malicious hackers (i.e black hat). They are the greatest concern, danger and threat for the black-hat community. This is due to them relying on the offence rather than the defence. Additionally, they use aggressive methods, attacks and techniques to take down them down. This is done by either uploading viruses and destroying their devices and computers, or retrieving their information and killing off their devices~\cite{levy1984hackers}.
\item \textbf{Script Kiddie:}
exclusively rely on copying codes and scripts to perform  malicious code injection (i.e SQLi) or and modification, mostly. Moreover, Script Kiddies rely on already-made programs in order to use them. In some cases, script kiddies watch tutorial videos in order to apply them and use them themselves. In fact, they rely on Metasploit (but not always) to create a package or a virus. Script kiddies also focus on the use of DoS and D-DoS attacks to hit the availability of a  given system, website or device~\cite{kargl2001protecting}.

\item \textbf{Blue Hat:}
 hackers are classified as Script Kiddies who are trying to take revenge or vengeance against someone or something, using hacking as a tool. However, unlike green-hat, they have no desire to learn new tricks and techniques and rely on the knowledge that they have already gained to perform their tricks. However, in most cases these are old tricks and already known~\cite{levy1984hackers}. In fact, blue hat hackers can also be hackers from outside the computer security consulting firms that test a given system for any possible software vulnerability or a bug~\cite{fried2005blue}. They are classified as invited security professionals to find any vulnerability in Microsoft Windows.
\end{itemize}

\subsection{Classifying Hackers}
After classifying hackers into different categories, it is important to identify them and know the difference between them. This includes identifying their aims, objectives and goals. Therefore, this section is dedicated to highlight on the most known type of infamous hackers that already exist as follows.
\begin{itemize}
\item \textbf{State/Nation Sponsored:}
This type of hackers is employed by a given government with open and even unlimited resources available at their disposal, in addition to the employment and use of very advanced sophisticated tools. This helps them exploit any security gap or vulnerability to gain confidential information about another rival government to stay on the lead~\cite{jordan1998sociology}. Moreover, this also allows them to achieve further goals based on influencing a government election, like when Russia was accused of interfering in the US elections in 2016~\cite{ohlin2016did}.

\item \textbf{Hacktivist:}
Hacktivism usually occurs when a hacktivist mainly hacks for either a social or political agenda, mainly to protest against a government's actions or military actions~\cite{mansfield2011hacktivism,caldwell2015hacktivism}. They are based on an anonymous, or even a covert group which can either be local, regional or/and international. Such type of hackers rely on targeting and hacking a government, a military website, or/and an organisation in order to grab some media attention to propagandise their own objectives and goals~\cite{denning2001activism}. Their favourite attack types are (but not limited to) Denial of Service and Web Defacement.

\item \textbf{Whistle-Blower:}
is a malicious insider that can be hired by rival groups or organisations in order to cause internal damage to the system or/and its components, either logically or physically (cyber-physically), or even both. Such move is accomplished, based on their ability to easily gain authorised access, along with  performing unauthorised tasks without the organisation’s knowledge (i.e abuse of privilege). Therefore, disclosing business secret trades and targeting an organisation's reputation, business trade and customers. Thus, leading to devastating financial and economic losses~\cite{jordan1998sociology}.
\item \textbf{Cyber-Spy:}
Hackers might also be employed by a given ntional/international intelligence agency in order to perform cyber-spying, cyber-sabotage, cyber-espionage, cyber-warfare~\cite{janczewski2007cyber} or cyber-intelligence task(s)~\cite{borum2015strategic,goel2011cyberwarfare,libicki2017coming,morag2014cybercrime,matsubara2014countering} against a given cyber-government, cyber-military or cyber-terrorist target(s). It is mainly done to conduct cyber-espionage operation(s), or a covert information gathering to retrieve essential information about their targets for further exploitation and further cyber-attacks (i.e Reconnaissance)~\cite{nakashima2011report}. In fact, they can also be employed by rival organisations to conduct industrial cyber-espionage to ensure industrial data theft.

\item \textbf{Cyber-Heist:}
Cyber-heist is usually achieved when the attackers try to steal as much money as possible from a large number of known online bank accounts, in the least amount of time. Such attacks are usually achieved either through hacking or even phishing, where they are cashed out in a single unique operation, targeting either the customer of a given bank, or the bank itself~\cite{sablik2017cyberattacks}. A prime example of that is the Bangladesh bank cyber-heist~\cite{karim2016cyber}, and the cyber capabilities of North Korea~\cite{chanlett2017north}.

\item \textbf{Botnet Master:}
Such type of hackers eventually rely on targeting weakly secured or non-secured devices including computers, devices, laptops and tablets. Such exploitation, is usually achieved either through the exploitation of a security gap, or an existing vulnerability based on a software bug or failure, or even a misconfiguration. Once exploited, hackers can use these devices' resources to perform his attacks. This is done by turning these devices into bots~\cite{fabian2007my} to lead a DDoS attack, or other types of attack~\cite{verini2010great}.
\item \textbf{Cyber-Criminals:}
Hackers can also commit cyber-crimes, hence they are known as cyber-criminals~\cite{filshtinskiy2013cybercrime} when they try to perform and commit their attacks online. All of the crimes mentioned above, along with other crimes can be classified as cyber-criminal acts, and sometimes, cyber-terrorist acts~\cite{furnell1999computer}. This will be further discussed  in the next section, in more detail.

\item \textbf{Cyber-Terrorists:}
Cyber-terrorists also rely on the internet in order to lead and perform online attacks~\cite{denning2001activism,colarik2006cyber,lewis2002assessing} to cause serious economic implications~\cite{hua2013economic}. Such attacks can either be through the spread of cyber-attacks, based on web-defacement~\cite{alhamed2013website}, or even Denial of Service (DoS) and Distributed Denial of Service (DDoS) attacks. This is all due to the fact that terrorists’ knowledge is very limited in the cyber-world. However, it is also evolving using different methods, techniques and tactics~\cite{taylor2014digital,dawson2015understanding,charvat2009cyber}. A prime example of that is Al-Qaeda's cyber wing~\cite{choi2018spreading}, and ISIS's cyber-caliphate, cyber-recruiting, cyber-Jihad and e-terrorism~\cite{saad2015infowar,nance2017hacking,liang2017unveiling,mcelreath2018communicating}.

\item \textbf{Suicide Hackers:}
Suicide hackers are hackers who aim at hacking for a given purpose in order to achieve their intended objectives and goals~\cite{patil2017ethical,jones2011computer}. Suicide hackers do not care about suffering from long term imprisonment/jailing, nor being obliged to pay a huge fine as a consequence of their committed cyber-activity. Moreover, suicide hackers can either be surprisingly good hackers, or bad hackers.
\end{itemize}

\subsection{Hakers Motives}
It is also important to know the motives~\cite{lakhani2003hackers} behind a hacker’s intentions. This will help in understanding more any hacker’s background, and also establishing a better understanding about the motives that can lead a specific hacker to perform their cyber-attack.

\begin{itemize}

\item \textbf{Political:}
Political motives are usually based on targeting government’s official websites either through DoS and DDoS attacks, or through web defacement. This is done to register a protesting statement against the government’s new law, or a new agreement on military action against another government. Such moves can be usually led by anti-war hackers~\cite{vegh2013classifying} as a cyber-protest and demonstration.

\item \textbf{Patriotic:}
or national motives are based over using hacking as a method to register an official protest as part of condemning a given act against a given country. Hence, relying on hacking as a form of protest. In fact, it looks very similar to hacktivism. Except that the main difference is based on a given national group targeting a given country or an individual. The most famous example occurred when Chinese hackers did send series of DDoS attacks against American and NATO websites as a form of protest against the bombing  of the Chinese embassy in Belgrade, in 1999~\cite{gries2001tears,sweeney1999nato}.

\item \textbf{Religious:}
Religious motives are another reason for committing a cyber-crime in the name of a certain religion. Such attacks were formerly led by Al-Qaeda~\cite{gunaratna2002inside} and now ISIS/ISIL~\cite{farwell2014media,tinnes2018bibliography}, more specifically, ISIS cyber-wing (cyber-caliphate), to either recruit online, spread propaganda, or communicate with their sleeping cells in western countries to perform terrorist attacks~\cite{shehabat2017encrypted,atwan2015islamic}.

\item \textbf{Racism:}
Many hacking attacks were also led by racist groups including the KKK and Neo-Nazis, based on hacking with the purpose of ensuring a racial supremacy over the other races~\cite{daniels2009cyber}, or trolling~\cite{jakubowicz2017alt_right}. It can also be classified as a different form of terrorism led and performed in order to target other minorities relying on the spread of fear and terror, online~\cite{daniels2009cloaked}.

\item \textbf{Social:}
Social motives are usually led by hacktivists in order to lead cyber-attacks against racist groups, terrorists groups, and/or  governments, through hacking. Such cyber-attacks are conducted to make a clear statement of opposition towards groups or/and condemning their actions, through hacking. 
\end{itemize}

\subsection{Cyber-Activity}
Hacking can take part or even be part of a cyber-activity led by a hacking group or individual(s), either locally, regionally or globally. Even though, hacking can take many forms, depending on the motives mentioned earlier, and the cause that led hackers to use hacking. Such cyber-activity, can be cyber-crimes, cyber-terrorism, cyber-warfare, and cyber-espionage, targeting the IoT including Medical IoT~\cite{yaacoub2020securing}, Industrial IoT (Cyber-Physical Systems)
\cite{yaacoub2020cyber}, Military IoT (i.e Drones/UAVs~\cite{yaacoub2020security}).
\begin{itemize}
\item \textbf{Cyber-Crimes:}
Hackers mainly rely on cyber-theft in order to steal financial or personal information via computers~\cite{shariff2009confronting} or insecure browsers. This was done by breaching the security measures to intercept data including credit card numbers, pin codes and other sensitive information. In 2013, cyber-thieves stole credit and debit card information of around 40 million worldwide shoppers. Moreover, e-mail spamming abuses the electronic messaging systems by constantly sending empty messages~\cite{singh2012client}. This also includes the ability to target search engines, wikis and blogs~\cite{schwartz1998stopping}. It was adopted by the attackers to carry out various types of malware~\cite{zhang2012survey}. In 2005, two spammers violated the Florida Electronic Mail Communications Act~\cite{scanlan2005fight}. These spammers received a fine of 50,000 \$ with an additional charge if the spam was still on the spread. However, they evaded paying fines by providing a fake financial statement~\cite{zdziarski2005ending}. In May 2007, Robert A.S., was arrested and prosecuted for more than 35 criminal acts including identity theft, e-mail fraud, identity theft, and money laundering. In fact, cyber-heist is a large scale monetary theft~\cite{verini2010great} which was conducted by relying on digital devices to perform their cyber-crimes through hacking, with Crimeware kits~\cite{kotov2013anatomy} including SpyEye~\cite{protection200811}, Butterfly Bot~\cite{dupont2017bots} and Zeus~\cite{mohaisen2013unveiling} being the most used hacking tools. The most infamous cyber-heist acts occurred in December 2012, when Bank of Ras Al-Khaimah came under attack, loosing \$5 million~\cite{basamh2014overview}. 

\item \textbf{Cyber-Espionage:}
became another emerging field, especially when a Stuxnet worm attack targeted the Iranian nuclear power grids~\cite{farwell2011stuxnet}, followed by other similar attacks (i.e Flame, Duqu, Gauss)~\cite{bencsath2012cousins}. Another aspect of cyber-espionage. Moreover, industrial cyber-espionage~\cite{coskun2003counteracting} took place in 1997, where engineers disclosed sensitive business information to the company’s competitors. In fact, such attack was revealed through received emails and faxes. The engineer was accused of industrial espionage and was sentenced for around two and a half years in jail. A large spy network called GhostNet~\cite{deibert2009tracking} was revealed by Canadian researchers in 2009. It was responsible for more than a thousand computer intrusions in more than 103 countries, gaining unauthorised access, and compromising different devices. Chinese hackers were also accused of such cyber-espionage. Another case of cyber espionage infected the computers of both John McCain and Barack Obama during their presidential campaigns in 2008~\cite{sasso2013report}. In fact, Chinese and possibly Russian hackers did allegedly install a spyware on the computers of these two presidential candidates and stole sensitive data related to foreign policy. In 2009, the Pentagon reported that their multi-billion dollar project of the next generation Fighter-Jet was subjected to a cyber-attack from unknown intruders (possibly Chinese)~\cite{gorman2009computer}. This was achieved through a coordinated cyber espionage attack during a two-year period. Therefore, stealing a massive amount of data about electronics and internal maintenance. Operation Shady RAT is the most infamous cyber-espionage case that affected more than 70 companies and organizations since 2006~\cite{alperovitch2011revealed}. It was spread through an e-mail with a malicious link attached to it as a self-loading remote-access tool, or Remote Access Trojan (RAT). Thus, gaining authorized access to legal contracts, government secrets, and other sensitive data. Among the targeted organisations, International Olympic Committee, United Nations and World Anti-Doping Agency were also under attack.

\item \textbf{Cyber-Terrorism:}
Due to the availability of hacking tools and internet with high speed real-time data transmission, and the easiness to obtain laptops, computers, and mobile phones at low prices, it became a source of attraction for terrorists with psychological effects~\cite{gross2016psychological}. As a result, terrorists started using it as a new emerging type of cyber-warfare, known as the asymmetric cyber-warfare, in order to lead cyber-attacks against Western governments and military installations~\cite{bogdanoski2013cyber}. Their motives are usually either religious or racial, based on the spread of hate, terror, fear and/or racism through hacking. This was mainly done through defacing websites, which led to DoS and D-DoS attacks~\cite{albahar2019cyber}.

\item \textbf{Cyber-Warfare:}
or information dominance or/and information warfare~\cite{endsley1997situation}, which became a new  type of war, was achieved  through cyber-space against different countries~\cite{arquilla2012rebuttal}, in order to cripple their ability and smart infrastructure. A global cyber-war was already waged against ISIS in~\cite{liang2015cyber,nance2017hacking}. This war managed to decisively defeat their cyber-caliphate online, and freezing their online activities. Another attack was led by Russia against Georgia and Estonia in 2007-2008, by sending multiple waves of D-DoS attacks. Therefore, crippling their online available services~\cite{ottis2008analysis,traynor2007russia}. In fact, cyber-warfare can be part of cyber warfare and electronic warfare domain to ensure the Battlefield's Situational Awareness (BSA) using using Cyber Electronic Warfare (CEW)~\cite{yasar2012operational,askin2015cyber}.
\end{itemize}


\subsection{Hacking Gains}
This section is dedicated to classify them into either commercial gains, financial gains, personal gains, or even political. In fact, it depends on the motives of the intended hacking group, their background and ability to wage cyber-attacks.
\begin{itemize}

\item \textbf{Economic:}
Hackers do rely on targeting one’s economy for multiple reasons. Such reasons depend on their intended objectives and aims. In fact, they can be employed by rival organisations to target another competitive organisation and reveal its business secrets and trades. Moreover, they might also be employed to cripple the economy of a given organisation by hitting its online available services~\cite{cashell2004economic}. Furthermore, it seems like cyber crimes are responsible for around \$12 million per business annually~\cite{tuttle2017cybercrime}. 

\item \textbf{Financial:}
Financial gains are on top of priority of hackers. Moreover, they are a crucial pumping source of money theft and money laundering, of which hackers heavily rely on as an illegal source of income. The global cost of a serious cyber-attack can lead to \$ 120 billion of losses. According to the PwC’s Global State of Information Security Survey 2018~\cite{TheGloba89:online}, the total cost for a financial incident was 857,000 euros, which is quiet high. However, according to Ponemon’s 2017 Cost of Cyber Crime Study, it has been revealed that over the last five years, the average cost has risen to 62\% compared to 27.4\% in 2017.

\item \textbf{Personal:}
Personal gains are another aspect, especially when a rogue/unsatisfied employee is kicked out of a given organisation. In fact, their aim shifts towards taking revenge against his own organisation. Such a move is done, by selling secrets to rival organisations, or disclosing the organisation’s sensitive information~\cite{marsh2017ethical}. The main aim and goal of this move is to either damage their reputation, scam and blackmail, or disclose their personal information, which can be sold for malicious third parties, or rival organisations.

\item \textbf{Political:}
Political gains are also of a high priority for a given political party, which can operate within the same country, or even against a foreign country’s political law. Therefore, hacking can be the new weapon to affect voting polls, or even election polls. In fact, Russia was accused of interfering with the US elections in 2016~\cite{ohlin2016did}. Such move is aimed at influencing another country through a political alliance achieved by paid propaganda and media, which plays a crucial and essential role in affecting the public's opinion.
\end{itemize}

\subsection{Targeted Systems}
The new adaptation and the widespread use of the internet was based on the digitising of written data and records into digital data and information stored in the cloud services. This led hackers to finding it as a great way to lead their cyber-attacks against various yet numerous targets. These targets include banks,enterprises, governments, hospitals, individuals, and organisations.

\begin{itemize}
\item \textbf{Banks:}
The most common attacks against banks can be based on hitting the availability of a given bank and preventing legitimate users from accessing and using the available services. This can have devastating effects on systems and people alike, targeting and compromising both systems and data's confidentiality, integrity or/and availability. Moreover, cyber-heist~\cite{chadwick2001bank,lennon2015hackers} can also occur. This leads to stealing online money as much as possible, in the least amount of time.
\item \textbf{Enterprises:}
are also prone to cyber-attacks led by hackers to disrupt their services and interrupt them. In fact, smaller enterprises are the most enterprises vulnerable to attacks by hackers. By quantifying a given risk~\cite{garg2003quantifying}, financial losses managed to reach up to more than \$445 Billion yearly, which is a quite high number.

\item \textbf{Governments:}  are mainly targeted either by hacktivists, cyber-terrorists~\cite{bergal2017hacktivists}, or even cyber-espionage attacks. Hacktivists perform their attacks as a form of protest, relying on hacking as a tool in order to make a clear statement. However, cyber-espionage is led by foreign rival governments against other governments by relying on cyber-intelligence agencies. In fact, it is a part of information gathering in order to lead a much more sophisticated cyber-attacks against smart infrastructure or even affect election polls~\cite{herzog2011revisiting,nakashima2016russian}.

\item \textbf{Hospitals:}
also became the perfect target for hackers in order to steal the patients’ private data and sell them to malicious parties. In  May 2017, the National Health Service (NHS) came under a WannaCry ransomware attack, led by North Korea's Cyber-Unit 180 (Lazarus)~\cite{maron2017us}, before establishing a security operations centre and investing £\ 250,000 to raise awareness and train NHS employees. In January 11th, 2018, Hancock Regional Hospital in Indianapolis was infected by a malware via an e-mail, known as phishing attack~\cite{maron2017us}. Hackers locked the hospital’s computer systems and demanded a ransom in Bitcoin crypto-currency. Therefore, luring the hospital to pay \$55,000 as a ransom~\cite{coventry2018cybersecurity}. 

\item \textbf{Individuals:}
can also be targeted. In fact, it depends on the intended target. More precisely, targeting the organisation, or bank, or even the enterprise’s employees, through social engineering, reverse engineering and phishing attack types~\cite{irani2011reverse}. Such targeting is based on the exploitation of these employees who are poorly trained or unsatisfied, by exploiting their “human feelings” to achieve a hacker’s goal(s).

\item \textbf{Military Websites:}
A new emerging technique is usually led through two main aspects, cyber-warfare~\cite{drahoscyberwarfare,stinissen2015legal}, or cyber-terrorism~\cite{conway2002cyberterrorism}. It can also take a secondary aspect by being part of a hacktivism campaign led by protesters (i.e anonymous) against a given military action~\cite{vinton2014syrian,denning2001activism}. Therefore, the military seems to be a recent cyber-target by different hackers~\cite{geers2014world}, where bits and bytes are replacing bullets and bombs in the cyber-world.

\item \textbf{Organisations:}
from any type are also prone to various cyber-attacks~\cite{gish2013effects,wrightson2014advanced}. These cyber attacks are based on targeting the organisations’ devices, relying on phishing and spear-phishing attacks~\cite{caputo2014going,hong2012state}. These attacks can infect a given organisation device with a malware capable of recording keystrokes, or injecting a spyware, or a Trojan horse (RAT). This turns the devices into botnets, or gain advanced access privileges. 
\end{itemize}

\begin{figure*}[!ht]
  \centering
  \begin{minipage}[b]{0.9\textwidth}
    \includegraphics[width=\textwidth]{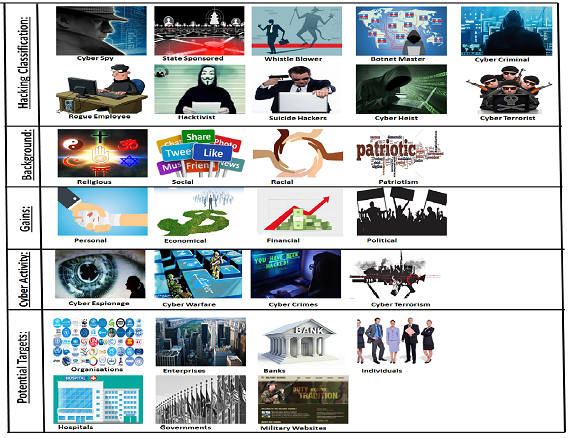}
    \caption{Cyber-Classification}
    \label{fig:ca}
  \end{minipage}
\end{figure*}

Finally, this paper summarizes them in \figurename~\ref{fig:ca}.
\label{sec:1}

\section{Vulnerability Assessment \& Risk Specification}
\label{sec:2}
Untested systems tend to be more prone and vulnerable to various attack types. This imposes a higher risk of them being exclusively targeted due to their exploitable weaknesses and present security gaps. Therefore, it is important to assess vulnerabilities before specifying the risk(s) surrounding it.

\subsection{Vulnerability Assessment}
\begin{figure*}[!ht]
  \centering
  \begin{minipage}[b]{0.75\textwidth}
    \includegraphics[width=\textwidth]{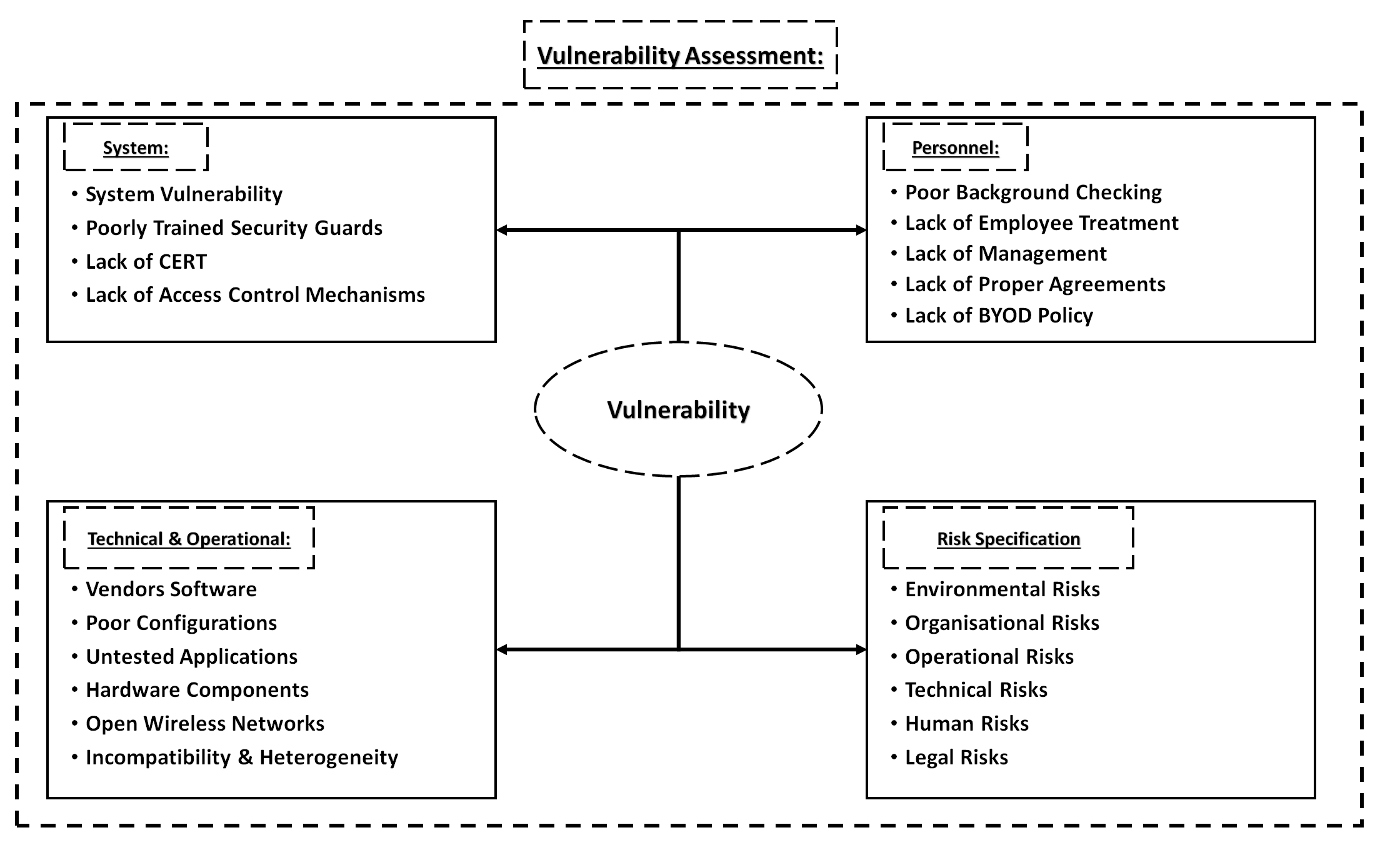}
    \caption{Pen Testing Appliance}
    \label{fig:1}
  \end{minipage}
\end{figure*}

In order to evaluate a given vulnerability, it is highly recommended to assess it first before taking any further step. In fact, the first step towards assessing a given vulnerability, is to identify it. In~\cite{votipka2018hackers}, Votipka et al. were among the first to conduct an analysis comparison and study between ethical hackers and testers. The analysis included their way to scan, find and discover vulnerabilities. Furthermore, their paper included how testers and hackers develop their own skills, knowledge and experience, as well as the tools in use. Moreover, the authors also provided their own recommendations to support and improve security training for testers, along with better communication between ethical hackers and developers, where hackers were also encouraged to participate. \\ 
In fact, it is classified in \figurename~\ref{fig:1}, and is described as follows.

In fact, there is a huge number of Vulnerability Assessment Penetration Testing (VAPT) tools~\cite{goel2015vulnerability} that can be used in order to perform a pen testing and vulnerability assessment for ethical hacking purposes. Different papers mentioned various types of vulnerability assessment and penetration testing tool such as in~\cite{goel2015vulnerability,antunes2009comparing,arkin2005software,de2016pentesting}. However, the main VAPT tools in use will also be highlighted. This is due to them being used during the ethical hacking training among other tools. Furthermore, \tablename~\ref{tab:4} presents a summary of the most frequently used tools.
\begin{table*}[!ht]
\centering
\small
\caption{\textbf{VAPT Tools }}
\label{tab:4}
\begin{tabular}{|p{4cm}|p{12cm}|}
\hline
\textbf{VAPT Tools} & \textbf{Description} \\ \hline
 Nessus  & Vulnerability Scanner \\ \hline
Metasploit & Vulnerability scanning and exploitation \\ \hline
 Kali Linux & A various collection of different tools \\ \hline
Burp Suite  & scans for web vulnerabilities \\ \hline
Paros Proxy  & scans for web vulnerabilities \\ \hline
Core Impact  & vulnerability exploitation and scanning \\ \hline
Nexpose & vulnerability scanner, supports the entire vulnerability management lifecycle (detection, verification etc..)\\ \hline
BeEF (Browser Exploitation Framework)& assesses the actual security posture of a targeted environment through a client-side attack \\ \hline
THC-Hydra & network login cracker that supports different services \\ \hline
Aircrack-ng & Assesses wireless networks, covers, captures and attack packets (cracking WEP, WAP, WPA, etc..) \\ \hline
Parrot Security OS & designed for vulnerability assessment and mitigation penetration testing,, computer forensics and anonymous web browsing \\ \hline
\end{tabular}
\end{table*}

\subsubsection{System Vulnerability}
System vulnerabilities can take many different aspects and actions if the right security measures were not taken into consideration and implemented on time. Otherwise, the system will be vulnerable to various types of threats and attacks. This can have serious implications and consequences against a given system or installation.

\begin{itemize}

\item \textbf{Poorly Trained Security Guards:}
In many cases, the lack of training that the security guards undergo is scarily dangerous. For example, an insider masqueraded as an employee, cleaner or client can overpass all the security guards if they have good manipulative social skills. This allows him to gain an unauthorised access, steal essential documents, add an infected USB, gain a rootkit privilege and leave the organisation without the guard’s knowledge. Therefore, it is essential to keep training constantly and testing security guards to evaluate their response level, performance, alertness, readiness and awareness.

\item \textbf{Lack of Awareness:}
In many cases, employees do not turn off nor lock their devices when they are away from their desk. As a result, confidential information written on papers can be stolen by employees without anyone’s knowledge. Moreover, when the company throws its own papers in trash and garbage, hackers rely on it as a source of treasure to study and exploit them. Hence, the need for a much more effective training and awareness levels is essential to be applied in a given organisation. This includes destroying any sort of evidence before having it thrown. This also includes destroying hardware equipment and the use of anti-forensics tools and techniques to leave no traceable data that serves as evidence against a given company.

\item \textbf{Lack of CERT:}
Described as the lack of training that the IT security staff undergoes in order to become part of the “Computer Emergency Response Team” (CERT)~\cite{houle2001trends}. In fact, it is highly recommended to address to such a vulnerability and threat accurately, effectively, efficiently, and on time. 

\item \textbf{Lack of Access Control Mechanisms:}
The lack of the right access control mechanisms can make the system prone to various types of attacks and vulnerabilities. In many cases, employees are given access privileges to perform more than their intended task(s), along with being left with these privileges even after their task(s) have been performed. However, in other cases, physical access control is  another concern, since locks can be easily broken and keys can be easily obtained. Therefore, the adoption of biometric access control mechanisms~\cite{al2013biometrics,jain2012biometric,woodward2003biometrics} is important and deserves a new aspect of attention, especially since many biometric features can be imitated~\cite{douglas2018overview}.

\end{itemize}

\subsubsection{Personnel Vulnerability}
Personnel vulnerability is based on the ability to hire, deal, manage and separate the right candidates from the wrong ones needed in a given domain. The wrong choice of personnel can lead to a possible insider attack, where a given employee may be employed by a rival organisation to perform such a malicious act. Therefore, security screening and background check is a must. However, personnel vulnerabilities are classified as the following.

\begin{itemize}

\item \textbf{Poor Background Checking:}
Candidates that are applying to a given job, may lie in their Curriculum Vitae (CV). In in many cases, these resumes are either over-exaggerated, or misleading. In fact, candidates lie about their experience, knowledge and skills in order to gain a given job offer, with fake references being added and fake recommendation letters and qualifications being presented. In some real-life scenarios, people with a previous legal convictions might not use or add it to their CV, in order to evade being detected or hide this part of their identity and aspect. Therefore, it is highly recommended to have an overall background check over any accepted employee by testing them and conduction training operations to check and assess their level of experience, knowledge and skills. This also includes conducting a background security check and clearance before being given a job. As a result, the need for employee screening and background check must always be fulfilled.

\item \textbf{Lack of Employee Treatment:}
When employees are poorly/ill-treated,  their work and productivity are affected. This would have a negative impact in return towards a given organisation’s profit, protection and productivity. Such lack of treatment could be based on paying low wages, or employee’s dissatisfaction. This leads to the possibility of having employees being exploited by a malicious third party through social engineering, phishing or even rogue employee recruitment. This later on can lead to lying, disclosing, leaking and stealing confidential and sensitive credentials and information. Therefore, it is highly important to keep all employees satisfied in every possible way, to maintain and increase their productivity, protection and profit levels.

\item \textbf{Lack of Management:}
Lack of management of employees without the necessarily required credentials based on their levels of knowledge and experience can have negative effects. Moreover, the poor management of employees is to not employ them in the right domain and the right place. Therefore, employing them in the wrong spot, wrong place based on political intervention, or religious intervention like in the Middle Eastern countries can have a serious and negative effect and impact on all the aspects of a given organisation. This would lead to poor and inconsistent security measures, lack of leadership skills, lack of team spirit, and lack of cooperation and collaboration among the different peers of the same given group.

\item \textbf{Lack of Proper Agreements:}
The lack or misuses of agreements between the company and its employees is also of a problem. This is based on the nature of the agreement signed by the employee and his commitment to the given organisation. In many cases, the agreement is violated due to the lack of order, misunderstanding and/or ill discipline. Therefore, it is essential to identify three main types of agreements including:
\begin{itemize}
    \item \textbf{NDA:} also known as Non-Disclosure Agreement. This agreement states that no employee should reveal any sensitive information that may damage the organisation's reputation and line of work and business. Otherwise, the employee will be legally sued, penalised and punished.
    \item \textbf{NCA:} also known Non-Competent Agreement. The agreement states that upon the leave of a given employee from their previous company, they are not authorised to work in another competent company nor organisation, until after a given period of time (2-3 years if not more) This is due to the fear of the previous organisation from using his gained skills and experience in his new job.
    \item \textbf{SLA:} Also known as Service Level Agreement. This agreement states that a certain quality of service level must be maintained between the service vendor and service provider. Moreover, a special focus must be aimed at how to maintain a strong security level.
\end{itemize}
\item \textbf{Lack of BYOD Policy:}
The lack of the implementation of the Bring Your Own Device (BYOD) policy can have its own serious implications. In fact, it offers the opportunity for any rogue employee to conduct a remote/insider attacks from within the organisation itself, without any knowledge about the source of the attack. In other cases, any employee's misuse of their device can lead to the exploitation of other devices. This can be done by simply plugging in his device to recharge it from a nearby laptop, or wants to upload a photo on his own working desktop device. Moreover, this rogue employee can use the victim's infected device without their knowledge that their device is infected. This can lead to the spread of a worm malware attack which can replicate and affect all the nearby devices.
\end{itemize}

\subsubsection{Technical \& Operational Vulnerability}
Operational vulnerabilities can be divided between software bugs and misconfigurations, as well as firmware bugs and misconfigurations, along with the application’s implementation issues. In fact, an operational vulnerability is based on the physical hardware equipment and components being employed and added to a given system. Every single aspect of technical and operational aspect is prone to various types of vulnerabilities including the following:
\begin{itemize}

\item \textbf{Vendors Software:}
In many cases, vendors employ software without taking into consideration the security aspect of its given design. Moreover, software is either employed with no security protection, or with very weak security protection. This leads to compromising the software which will be prone to various types of attacks. Thus, leaving the system vulnerable to various attack types. Therefore, it is essential to ensure a software pen testing as well~\cite{arkin2005software,potter2004software}.

\item \textbf{Poor Configurations:}
In many cases, many software types are prone to various misconfigurations. It is due to the fact that they are poorly and hastily configured, without taking into consideration the importance of testing each software’s configuration. Thus, leaving it non-secure against any possible form of malicious code injection attack type. Therefore, there's a higher need to employ coding experts, capable of writing and checking a given software code and testing it, before making it available for the market use.

\item \textbf{Untested Applications:}
Many applications remain untested and unevaluated to check their security and vulnerability levels. Therefore, these applications are left unevaluated without being deemed as either secure or non-secure. However, if they are deemed as secure, these application can be safely used. If deemed as not secure, these applications can be prone to various attacks types including surveillance attacks, which are based on masquerading a malicious application as a normal application. Therefore, it is essential to test and verify each application before employing it, especially since applications are becoming a commercial competitive way between different peers. This will avoid focusing on the importance of evaluating the levels of security of each application.

\item \textbf{Hardware Components:}
Many hardware components lack of the necessary physical protection and security against any possible physical attack based on an unauthorised access. Therefore, it is essential to protect servers and storage areas from any possible and unwanted physical access to prevent any possible attempt to break into the system and cause physical/logical damage and harm. In fact, it works in parallel with the employment of physical access control mechanisms to ensure a better protection.

\item \textbf{Open Wireless Networks:}
The communication between different peers and devices relies on open wireless networks. As a result, such devices are prone to eavesdropping, sniffing, replay and man-in-the-middle attacks, along with various different attack types that can intercept and alter transmitted information. This also includes hijacking and manipulating it without both source and destination’s knowledge. Moreover, some wireless communication networks are poorly configured and can be easily broken using password cracking methods and attacks.\\
Therefore, it is highly recommended to assess the level of vulnerability through scanning, which is beneficial for cyber-security experts and researchers alike~\cite{wang2017ethical}. It is also recommended to understand the vulnerability and the way it is exploited. In~\cite{trabelsi2018teaching}, Trabelsi et al. offered a way to teach information security students through an information security education over how key-logging is achieved along with the appliance of network eavesdropping attacks, as part of their ethical hacking program. Moreover, it is also important to assess a given vulnerability. This allows an easier identification and detection of the Advanced Persistent Threat (APT) strategy, especially since many attacking techniques seems to be similarly the same~\cite{tankard2011advanced}. Indeed,  \tablename~\ref{tab:8}  summarises the main network vulnerability scanning tools along their description. 

\begin{table*}[!ht]
\centering
\small
\caption{\textbf{Network Vulnerability Scanning Tools}}
\label{tab:8}
\begin{tabular}{|p{4cm}|p{12cm}|}
\hline
\textbf{Network Vulnerability Scan Tools} & \textbf{Description} \\ \hline
Microsoft Baseline Security Analyzer (MBSA) & identify any missing service packs, security patches, and security misconfigurations, scans one or more IP addresses, scan weak passwords, or SQL administrative vulnerabilities \\ \hline
Nexpose Community Edition &  scans network vulnerabilities, web applications, and virtual environments, limited scan of 32 IP (max) addresses simultaneously, infeasible for scanning a large networks \\ \hline
Nessus & provides vulnerability scanning for network devices, operating systems, databases, web applications, and IPv4/IPv6 hybrid networks alike \\ \hline
Nmap & flexible tool, capable of mapping network filled with packet filters, routers, and firewalls, scans large networks (thousands of hosts) and small networks (single host), found in many systems including Kali Linux ~\cite{alqahtani2013tcp} \\ \hline
OpenVAS & offers a comprehensive and powerful vulnerability scanning and vulnerability management solution, one of the most powerful free security scanners that you can use for free, offers false positive management of scanning results \\ \hline
Retina CS Community & find network vulnerabilities, configuration issues, and missing patches, provides free scanning and patching for 256 IPs (max), supports vulnerability scanning in mobile devices, servers, and web applications \\ \hline
\end{tabular}
\end{table*}

\item \textbf{Incompatibility \& Heterogeneity:}
 Different security mechanisms and methods became incompatible, especially with the constant updates and new devices being manufactured relying on the commercial aspect of competitiveness. This is due to the heterogeneity of the used and employed devices, which have  different software, firmware, and hardware components. However, manufacturers took a good grasp of the importance of security measures, and started working on deploying different type of security to protect genuine devices, unlike the cheap counterfeit non-secure devices.
\end{itemize}

\subsection{Risk Specification}
After assessing the  vulnerabilities according to both, the likelihood and impact of a given risk, it is important to specify and classify each given risk. Once achieved, risks must be mitigated by taking the right security measures and counter-measures in order to adhere, identify and overcome any possible risk from occurring. In~\cite{turpe2009testing}, Turpe et al. classified risks as technical, organisational and legal. Even though there is an agreement over this specification, except there are more risks including human risks, environmental risks and operational risks that should also be taken into consideration. 
\begin{itemize}

\item \textbf{Environmental Risks:}
 are based on the likelihood or the possibility of an earthquake from occurring. This also includes a tornado, hurricane, volcano, tsunami, flooding etc. Environmental risks must also be taken into consideration, especially when aiming to place a given server, or a backup server away from such locations in order to avoid and limit further physical damage and financial losses.

\item \textbf{Organisational Risks:}
 are usually caused by pen testers or employees by accident. This leads to unnecessary triggering incident responses and security measures, which in turn leads to resource exhaustion, and disruption of services. Moreover, in many cases, after performing the pen test, no focused attention is aimed at real case scenarios that might occur after a given test. This can have a real and negative effect on a pen testing team’s reputation. 

\item \textbf{Operational Risks:}
 are primary related to the malfunction of a given hardware component in a given device or system. This leads to a cascading error, and fatal crash of the system. Operational risks, are also based on the possibility of a power shortage, or higher level of electricity power that can destroy a given device or affect its availability. This can also disrupt, or even interrupt other available devices and services' availability. In fact, devices' over-usage can lead to the device overheating, which can damage its components and systems, and later on result into a sudden device stoppage.  

\item \textbf{Technical Risks:}
 are based on the possibility of a system or/and software failure to connect to other available systems or/and services. Such disruption can be deliberate through technical exposure led by rogue employees, or non-deliberate (done by mistake). Either ways, the service disruption or performance reduction can also be caused by either an occurring fire in a given room, or flood with water reaching servers and devices. This leads to either the loss, disruption, alteration, modification or/and disclosure of data. Data disruption can also be caused by ethical hackers and pen testers, which usually occurs by mistake, causing some serious fatal errors.

\item \textbf{Human Risks:}
Human risks should be also highlighted as a serious risk. For example, when performing pen testing, ethical hackers will play the role of rogue employees and insiders relying on social engineering and phishing techniques, tools and methods. Once inside, the simulated attack will be based on injecting malicious malwares, or plugging a keystroke USB, or gain access through a malware injection, or stealing confidential documents. The risk is based on either having the simulated attacker caught, or trespassing the organisation’s physical security checks and access controls. However, in many cases, the organisation is not notified about such a simulated attack, and can prosecute the ethical hackers for trespassing.

\item \textbf{Legal Risks:}
Legal risks occur when both sides do not sign a non-disclosure agreement, of which it is a two-edged sword risk. From the organisation’s part, ethical hackers can be rogue hackers that study the organisation and conduct fake pen testing for further exploitation through spotting vulnerabilities and security gaps without reporting them. Therefore, making it more prone to attacks after conducting the information gathering phase through a fake pen testing. As from the ethical hacker’s part, conducting simulated attacks without a legal approval can be classified as a criminal act by law. This can have them being legally prosecuted and punished for their committed acts.
\end{itemize}

\section{Targeted Cyber-Attacks}
\label{sec:3}
\begin{figure*}[!ht]
  \centering
  
    \includegraphics[scale=0.45]{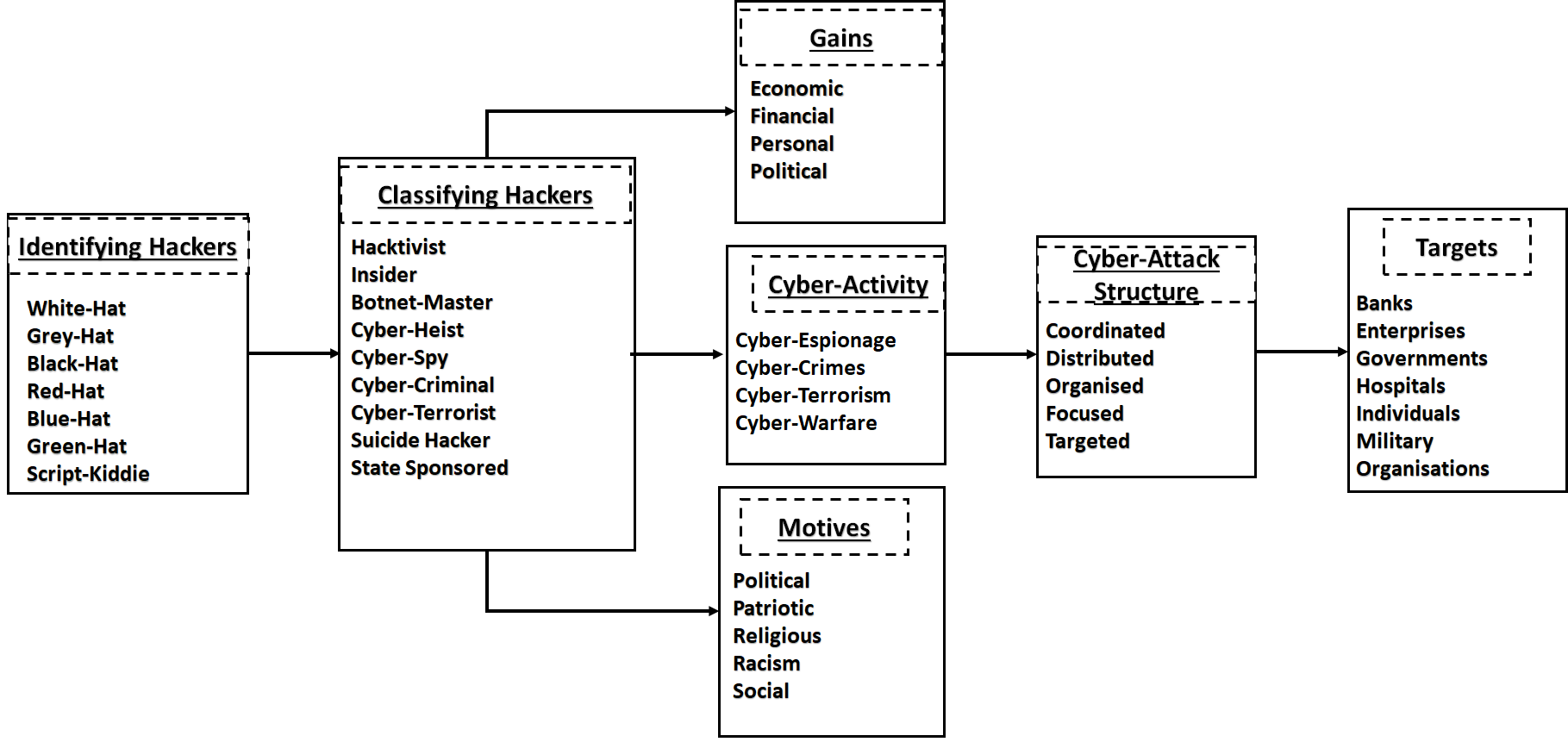}
    \caption{Proposed Identification \& Classification Framework}
    \label{fig:a}
 
\end{figure*}

Targeted attacks are presented according to their aim to target a given organisation, website or application (see \figurename~\ref{fig:a}). As a result, these attacks are presented as follows. However, before proceeding any further its important to identify the cyber-attack's structure. 

\subsection{Cyber-Attack Structure}
Cyber-attack structures eventually differ depending on the attack type, motives, knowledge, experience and skills of the attacker. Moreover, the attack structure also depends on the available resources at the hackers’ disposal. As a result, the cyber-attack structure is divided as follows: coordinated, distributed, organised, focused and targeted.
\begin{itemize}
\item \textbf{Coordinated Attacks:}
 occur through the synchronisation between different hackers located locally, regionally, or/and even globally to lead and perform their cyber-attacks simultaneously against a given target. Such type of attacks are usually very well-fit for leading and launching DoS and DDoS attacks to target the availability of a given website or system.
 
\item \textbf{Distributed Attacks:}
 are close to coordinated attacks, of which they are another form of coordination. Here, attackers lead a synchronous continuous attack against different distributed targets, which are located in different geographical locations. In addition, distributed attacks can be led by attackers that are located at different distributed areas to perform their cyber-attacks. Thus, making it difficult to track them down, stop and overcome them at once.
 
\item \textbf{Organised Attacks:}
 usually take place when the aim, objective and goal of the attackers meet. Such organised attacks are based on the collaboration and cooperation between hackers to perform their cyber-attack, swiftly. Such attacks are performed in a high level of accuracy, where hackers aim to steal as much information/money as possible in the shortest amount of time.
 
\item \textbf{Focused Attacks:}
 are based on having a clear scope and vision over the intended target. Hackers tend to focus their attack on a given organisation. As mentioned above, it is related to their motives that reveal how each hacking group aims to focus, lead and launch their cyber-attack on a given target.
 
\item \textbf{Targeted Attacks:}
are not limited to hackers who launch a cyber-attack against a given target. In fact, targeted attacks can be based on leading simultaneous attacks against different targets. The aim is to reveal and expose the hackers ability to lead simultaneous attacks against different worldwide targets that are located at different geographical locations.
\end{itemize}

\subsection{Cyber-Attacks Type}
Cyber-attacks are not limited to one aspect and classification. In fact, there are endless classifications with countless attacks. The aim of this paper is to present the most common known types of attack that hackers and criminals use, while also presenting suitable security measures to mitigate and overcome them.

\subsubsection{Web Attacks}
Web attacks are primary classified as attacks that target websites, or use websites as a launching platform for their remote cyber-attacks. Among these attacks, their main types are presented as follows:
\begin{itemize}
\item \textbf{Cookies Hijack:} attack steals the magic cookie that authenticates a user to the remote server. HTTP cookies are also used to a multi-website session, due to the easiness of stealing them from their victim's devices~\cite{sivakorn2016cracked} . 

\item \textbf{Sessions Hijack:}
also known as session hijacking or key—to-gain unauthorized access to device's information session~\cite{kolvsek2002session}. It usually occur when the "Pass-the-Cookie" technique is used~\cite{mukkara2013secure} to perform "blind" hijacking or man-in-the-middle attacks using sniffing programs.

\item \textbf{Phishing:} is a social engineering technique that rely on fraudulent attempt(s) to obtain sensitive information (E.g usernames, passwords, credits cards, secrets trades etc..) through social phishing~\cite{jagatic2007social},  whaling (targeting high-profile employees such as Chief Executive Officers, Chief O Officers and Chief Financial Officers)~\cite{hong2012state}, or vishing attacks (stealing users credentials)~\cite{griffin2008vishing}. In this attack, users are lured to access fake websites or through email spoofing. To overcome this attack user training and awareness are a must.

\item \textbf{Masquerade Attacks:} usually target connected networks and devices with weak authorisation by using a fake (network) identity to gain unauthorized access to personal information (E.g usernames, passwords, logons, logins etc..) through legitimate access~\cite{stolfo2014methods}. This attack can be performed by insiders (whistle-blowers within the organisation) or outsiders (remotely).

\item \textbf{Buffer Overflow:} or buffer overrun is a program anomaly that overruns and overwrites a buffer's boundary  with adjacent memory locations~\cite{cowan2000buffer}. C and C++, Java and Python are common programming languages associated with buffer overflows. Such attack would result into overwriting adjacent data, overwriting executable code(s), memory access errors, incorrect results, and constant crashing. Cryptographic and non-cryptographic solutions were presented to protect against buffer overflow attacks in~\cite{piromsopa2011survey}. Bound checking is a suitable solutions, but requires additional processing time. Moreover, randomizing memory layout and deliberately leave buffer space are other suitable solutions to overcome buffer overflow Operating Systems (OS) attacks.

\item \textbf{Website Spoofing:} includes creating fake (hoax) websites that mislead readers into believing its legitimacy by masquerading it as a legitimate website through a very similar design including similar Uniform Resource Locator (URL) or Cloaked URL, and a shadow copy of the World Wide Web~\cite{grossman2007whitehat}. Website spoofing is always associated with phishing or e-mail spoofing.

\item \textbf{Malicious Code Injection Attacks:}
 usually occur on poorly configured and poorly secured web applications~\cite{ray2012defining}. In fact, they can be easily exploited by any hacker through a malicious code injection or an SQL injection attack~\cite{mitropoulos2017fatal}. As a result, this malicious script will be running on the victim's device without their knowledge~\cite{jin2014code}. Therefore, being capable of retrieving all of the users credentials including emails, passwords, credit card numbers, along with cookies, browsers, and the device's/user's ID.

\item \textbf{SQL Injection Attacks:}
The impact of SQL injection attacks varies depending on the objective and aim of a given hacker~\cite{kumar2012survey}. Such an aim is based on either manipulating database information, gathering sensitive data, or even executing a DoS attack. In~\cite{nagpal2015tool}, SQL injections were divided into three different categories. These categories include three order attacks. The first order attack, is based on the attacker entering a malicious string to modify the executed code. Second order attack, is based on the injection into a storage executed by another action. Third order attack is the later injection, where the attacker manipulates the function by changing its environment variables. On the other hand, there are different SQL injection methods that can be applied~\cite{kindy2011survey,kumar2012survey}, and they are presented in \tablename~\ref{tab:2}.
\begin{table*}[!ht]
\centering
\small
\caption{\textbf{Different SQL Injection Attacks}}
\label{tab:2}
\begin{tabular}{|l|l|}
\hline
SQL Injection Types & Description \\ \hline
Blind Injection &  logical conclusions are derived from the answer to a true/false question regarding the database\\ \hline
Logically Incorrect Queries & Using different error messages to retrieve information to exploit and inject a given database \\ \hline
Piggy-Backed Queries & malicious queries are additionally inserted into an already injected original query \\ \hline
Stored Procedure & executing database’s built-in functions using malicious SQL Injection scripts/codes \\ \hline
Tautology & SQL injection queries injected so so they always result in a true statement \\ \hline
Timing Attack & collecting information by noticing the database response time \\ \hline
 Union Query & malicious query joined with a safe query using UNION to get other table related information \\ \hline
\end{tabular}
\end{table*}

\item \textbf{Cross-Site Scripting:}
or XSS is a web application security vulnerability based on a malicious code injection by malicious web users. In fact, recent attacks were capable of leveraging session IDs by relying on JavaScripts functions on other websites to inject them into different pages. This included online banking, instant messaging and emails. Therefore, being capable of stealing a user’s session which in turn, can be used to steal stored web cookies. This allows the attacker to replicate a user's session on a different machine~\cite{held2010cross,al2013malware}. Another form of XSS is based on the attacker sending HTTP requests at his own will from his victim’s machine, or sending false counterfeited requests relying on the use of an especially sophisticated tool(s).
 
\item \textbf{Cross-Site Request Forgery:}
Cross-Site Request Forgery (CSRF) takes advantage of any possible lack of authorisation that is either absent or weakly employed in poorly designed web applications. The first step of performing this attack, is based on the hacker making the victims run the malicious script through clicking on malicious link without their knowledge. Therefore, allowing the attacker to use the victims' saved credentials to perform further attacks on their behalf, and without their knowledge. Such attacks usually take place through a network request injection relying on the browser of the user, which permits a website to show every HTTP request to any given network address~\cite{barth2008robust,al2013malware}.
\end{itemize}

\subsubsection{Application Attacks}
Application attacks can take many forms from covert attacks (surveillance) to much more overt ones such as viruses. This paper discusses them in a way that proves how they can be linked together.
\begin{itemize}

\item \textbf{Scareware:} uses social engineering techniques to trick their victims into downloading/purchasing unwanted dangerous rogue software (ransomware) by using fear as their tactic~\cite{seifert2015scareware}. This is caused by urging users to download a given (fake) Anti-Virus to remove a recently detected (fake) virus to remove it. Scareware can take many forms including antivirus, anti-spyware, firewall application or a registry cleaner. Scareware may include a clickjacking feature that redirects users to malicious website that triggers a malware download once the users clicks on the “Cancel” or the “X” buttons to close the window.

\item \textbf{Spyware:} is a software that installs itself on a given device to covertly monitor its victims' online/offline behaviour without their knowledge and permission~\cite{egele2007dynamic,hu2005spyware} and gain details about them (E.g names, addresses, browsing history, downloads etc..). Spyware can act as adware by marketing data firms, or can be a Trojan. Such attack can degrade a system's performance, damages the Central Processing Unit (CPU) capacity, affect the disk usage and cause network traffic overhead.

\item \textbf{Trojan:} is often disguised as legitimate software employed by cyber-thieves to gain access to users' systems~\cite{kang2004trojan}. Once activated, Trojans can steal, spy, exploit, implement backdoors (rootkit) to delete, block, modify, disrupt and modify data~\cite{tehranipoor2010survey}. Trojans can take many types Trojan-Banker (steal data's account), Trojan-DDoS (denying service), Trojan-Downloader (download/install new malicious program versions), Trojan-Dropper (install viruses), Trojan-FakeAV  (fake Anti-Virus), Trojan-GameThief (steals online gamers' accounts), Trojan-IM (steals users' usernames/passwords of Skype, Yahoo, Messenger etc..), Trojan-Ransom (modify computer's data as part of ransomware attack), Trojan-SMS (sending text messages from their victims' phones), Trojan-Spy (implement keystroke, keep users under surveillance), and Trojan-Mailfinder (harvest email addresses).

\item \textbf{Botnet:} bots or zombies exploits devices that suffer from the same vulnerability~\cite{stone2009your}, and can also be used to launch DoS and DDoS attacks~\cite{stone2009your,zhang2011survey} by being ordered from a command and control (C\&C) software. Botnets can steal data, send spam, or gain remote access. Botnets can take many forms including Hybrid Botnets~\cite{wang2008advanced,pieterse2013design}, Internet Relay Chat (IRC) botnets~\cite{lashkari2011irc}, Cloud Botnets~\cite{clark2011botclouds}, Peer-to-Peer (P2P) botnets~\cite{wang2010peer}, Hyper Text Transfer Protocol (HTTP) botnets, and Mobile (SMS/Bluetooth) botnets~\cite{liu2009botnet}.

\item \textbf{Virus:} is designed to spread from host to host~\cite{nachenberg1997computer} either through human interaction (Trojan, Logic Bomb~\cite{denning2000cyberterrorism,northcutt2005logic} etc..) or without human intervention (worm)~\cite{li2008survey}. Unlike worms, viruses cannot spread or reproduce without human interaction, since it is a malicious code or program written to alter the way a computer performs. Viruses can be attached to a program, file, funny images, audio/video files, socially shareable content or document, to steal data/passwords, register keystrokes, corrupt files, send spam, erase data or cause hard disk permanent damage. Infected devices show various symptoms including frequent pop-up windows, homepage changes, frequent crashes, and slow performance. Therefore, its always important to use trusted/verified up-to-date anti-viruses, avoid clicking on pop-up advertisements, and use spam filters.

\item \textbf{File Infection:} is a compute virus that inserts malicious codes into executable files on a given system~\cite{szor2005art}. They can overwrite various files or change their signatures to install the newly infected files and replace them with the original ones. This attack type can target various Operating Systems (OS) including UNIX, Macintosh, and Windows.

\item \textbf{Rootkit:} is used to gain a remote administrator-level access to a computer or network by stealing an administrator password or exploiting a system's (OS, firmware or application) vulnerability~\cite{chiang2007case}. Rootkits can deactivate/destroy anti-malware software to avoid detection and make its tracking extremely difficult. Moreover, rootkits can ensure a backdoor access through keylogging, or turn a given vulnerable device into a bot. 

\end{itemize}

\subsubsection{Network Attacks}
Poorly secure and non-secure networks are the most vulnerable to various threats and attacks that target them and exploit their security gap. As a result, the most commonly known type of these attacks is presented as follows:

\begin{itemize}

\item \textbf{Eavesdropping:} is known as a passive interception of communication between two parties and can take many forms such as: sniffing by installing a network monitoring software called sniffer, or snooping attack where an eavesdropper passively intercept non-secure or weak communication between two parties to steal their information and credentials via the communication network~\cite{zhang2004improving}. To overcome this attack, avoid public Wi-Fi networks, use firewalls, proxies and Virtual Private Networks (VPN).

\item \textbf{Replay:} are known as playback attacks that are a lower tier versions of a "Man-in-the-middle attack" where an attacker eavesdrops on a secure network communication and intercepts it for further repeated transmission later on to cause delays or disruption of service~\cite{syverson1994taxonomy}. To overcome this attack various solutions can be presented such as using timestamps, securing data storage with key update~\cite{mclellan2009secure}, using security protocols~\cite{malladi2002preventing}, using attentive filtering networks for audio replay attack detection~\cite{lai2019attentive}, and using Magnitude and Phase Information with Attention-based Adaptive Filters~\cite{liu2019replay}.


\item \textbf{Man-in-the-Middle:} (MITM) is an active eavesdropping form where the attacker impersonates each endpoint to alter, intercept and directly monitor the communications between two parties and tricks them into believing that they are communicating with each other~\cite{callegati2009man}. Therefore, encrypting communication is required and using a mutually trusted certificate authority between both parties is a must.

\item \textbf{Packet Sniffing:} intercepts the data through the capture of traffic using a sniffer software~\cite{thakur2013content} over non-secure channels and reads the unencrypted data. Captured data can be analysed to gain access or information. To overcome this issue, channels must be secure and data must be encrypted. 

\item \textbf{Password Cracking:} is a cryptanalysis process that aims to crack passwords to recover credentials and gain access to systems (System Administration privileges) and data~\cite{weir2009password}. Password cracking can range between brute force~\cite{owens2008study} (protected using multi-factor authentication), dictionary~\cite{narayanan2005fast} (protected using a passphrase), Meet-in-the-middle~\cite{demirci2008meet} (protected using stronger keys), online/offline password guessing~\cite{kelley2012guess} (protected using encrypted password form), rainbow-table~\cite{papantonakis2013fast} (protected using salting technique), and birthday~\cite{bellare2004hash} (protected using hashing) attacks with many tools being used and available for this task.

\item \textbf{Traffic Analysis:} intercepts and examines encrypted/unencrypted network traffic to recover any useful information (header, message length, repeated patterns, processing time, transmission delay, etc)~\cite{kadloor2010low}. This renders large, periodic, encrypted and plaintext traffic under constant watch. To overcome this attack, sophisticated cryptographic solutions are needed, as well as the use of The Onion Router (TOR).

\item \textbf{Wireless Jamming:} is used to compromise a secure/non-secure wireless environment (mainly Local Area Networks (LAN))~\cite{kumar2013jamming} by denying the access and transceiving services to authorized users, by blocking and jamming all wireless legitimate traffics on all targeted frequencies (mainly 2.4 GHz). To overcome this attack, Wireless LANs can be used, along frequency hopping and shifting techniques.

\item \textbf{Black-hole:} packet drop attack or blackhole attack is classed as a denial-of-service (DoS) attack by employing a router that (sometimes selectively) relays packets instead of discarding them~\cite{tamilselvan2007prevention}. Due to the lossy network nature, packet drop attacks are very hard to detect and prevent. Hence, they are frequently deployed to attack wireless ad-hoc neworks. This type of attacks can be mitigated using a Timer Based Baited Technique presented in~\cite{yasin2018detecting}.

\item \textbf{Byzantine:} attacks have total control on the number of authenticated devices before behaving arbitrarily~\cite{geetha2016byzantine}. This type of attacks turns insider nodes into malicious ones by preventing route establishments, modifying route selections, and dropping route requests. Therefore, disrupting and degrading the performance of both network and routing services. Byzantine takes many forms of attacks including Byzantine Wormhole attacks, and Byzantine Overlay Network Wormhole Attack, along others which are further discussed in~\cite{geetha2016byzantine}.

\item \textbf{DoS/DDoS:} a denial-of-service (DoS) attack prevents legitimate users from accessing their services through an excessive sending of authentication requests with invalid return addresses~\cite{wood2002denial}. A Distributed-Denial-of-Service (D-DoS) compromises multiple systems (compromised through a Trojan and turned into bots) before simultaneously flooding the victims with a massive incoming traffic from multiple sources. DDoS attacks have many types including: traffic attacks,  flooding Transmission Control Protocol (TCP) packets, flooding User Datagram Protocol
(UDP) and Internet Control Message Protocol (ICMP) packets, bandwidth attack (sending massive data junk), and application attacks (depleting application layer's resources).

\end{itemize}

\subsubsection{Infrastructure Attacks}
Infrastructure attacks are designed to target and cripple a nation's or competing organisation's infrastructure~\cite{nash2005undirected} such as the case of the American power grid incident on March 5th,  2019 (via firewall vulnerability)~\cite{weiss2019assessment,oughton2019stochastic}. Such attacks are part of cyber-industrial-espionage~\cite{albanie2019deep,heickero2019cyber} to cripple a nation's infrastructure by targeting control systems, energy resources, finance, telecommunications, transportation, and water facilities \cite{genge2015system}. Therefore, rendering Industrial Internet of Things under constant attack.

\begin{itemize}
\item \textbf{Insider Attack:} known as insider threat, is primarily led by malicious, rogue or unsatisfied employees with an authorized system access~\cite{salem2008survey}. They are the main vital link for outsiders to gain and perform cyber-attacks.

\item \textbf{Outsider Attack:} does not have a direct access to a device or network~\cite{walton2006balancing}. However, outsiders are linked to insiders who grant them a remote access to perform their attacks remotely.

\item \textbf{Key Stroke Register:} is known as keylogging or keyboard capturing~\cite{leijten2013keystroke} and can either be a software or hardware. It is based on covertly recording (logging) the keys struck (including key/microhphone register) on a keyboard to retrieve users' recorded data. Their primary use is to steal passwords and confidential information.

\item \textbf{USB Infection:} imposes a serious threat to critical infrastructure systems~\cite{anderson2010seven} since it can be used maliciously to deliver a malware to steal critical data and cause malicious attacks. A prime example of this attack was the Stuxnet worm in 2010 that targeted the Iranian power plants at Natanz, part of the allegedly joint American/Israeli cyber-espionage operation against Iran~\cite{denning2012stuxnet}. USB attacks can take many forms including a list of 29 different USB attack types that was presented in~\cite{nissim2017usb} at Ben Gurion University. These attacks can be categorised as Reprogrammable microcontroller USB attacks, Maliciously reprogrammed USB peripheral firmware attacks, Attacks based on unprogrammed USB devices, Electrical attacks used by USB killers, USB Switchblade, and USB Drive Infection.


\item \textbf{Social Engineering:} includes performing malicious activities through human interactions using psychological manipulation to trick their victims into revealing secret trades through the creating of fake trust, or through curiosity or fear~\cite{krombholz2015advanced}. To overcome this threat its important to avoid opening emails and attachments from suspicious sources, avoid tempting offers, keep anti-viruses up to date, while also using multi-factor authentication.

\item \textbf{Reverse Engineering:} includes attempts to steal sensitive data~\cite{chikofsky1990reverse} or cryptographic keys by relying on code tampering by introducing artificial bugs, as hackers are familiar with common structures in compiled code. In fact, they rely on code structures to locate and trace them, using sophisticated techniques such as static analysis~\cite{torrance2011state}. Recently, reverse engineering is now being used to hack connected cars~\cite{ring2015connected}. 

\end{itemize}

\par
Finally, a Pen Testing Risk assessment is presented in \tablename~\ref{tab:451} with its suitable security measure is presented per attack depending on which main security goal ((C)onfidentiality, (I)ntegrity, (Av)ailability and (Au)thentication) it targets.
\begin{sidewaystable*}[!tbp]
\centering
\caption{  \textbf{Pen Testing Risk Assessment Per Attack}}
\label{tab:451}
\centering
\footnotesize
\begin{tabular}{|p{3cm}|p{3cm}|p{1cm}|p{1cm}|p{1cm}|p{1cm}|p{1cm}|p{1cm}|p{3cm}|p{6cm}|}
\hline
\multicolumn{2}{|l|}{\textbf{Attack}}         & \multicolumn{4}{l|}{\textbf{Targeted Security}}                & \multicolumn{2}{l|}{\textbf{Exposure Risk}} & \multicolumn{2}{l|}{\textbf{Security Measures}} \\ \hline
\textbf{Target}                   & \textbf{Nature} & \textbf{C} & \textbf{I} & \textbf{Av} & \textbf{Au} & \textbf{Before Pen Testing}      & \textbf{After Pen Testing}     & \textbf{Cryptography}      & \textbf{Non-Cryptography}     \\ \hline
\multirow{10}{*}{\textbf{Web Attacks}} & \textbf{Cookies Hijack} & \textbf{\checkmark} & \textbf{X} & \textbf{X} & \textbf{\checkmark} & \textbf{High}      & \textbf{Low}     & \textbf{Encryption}      & \textbf{HTTPS, Secure websites}     \\ \cline{2-10} 
                            & \textbf{Session Hijack} & \textbf{\checkmark} & \textbf{\checkmark} & \textbf{X} & \textbf{\checkmark} & \textbf{High}      & \textbf{Low}     & \textbf{Encryption}      & \textbf{Stronger Authentication}     \\ \cline{2-10} 
                            & \textbf{Phishing} & \textbf{\checkmark} & \textbf{X} & \textbf{X} & \textbf{\checkmark} & \textbf{High}      & \textbf{Low}     & \textbf{-}      & \textbf{Training Awareness, Limited Privileges}     \\ \cline{2-10} 
                            & \textbf{Masquerade Attack} & \textbf{\checkmark} & \textbf{X} & \textbf{X} & \textbf{\checkmark} & \textbf{High}      & \textbf{Low}     & \textbf{Encryption}      & \textbf{Multi-Factor Authentication}     \\ \cline{2-10} 
                            & \textbf{Buffer Overflow} & \textbf{X} & \textbf{X} & \textbf{\checkmark} & \textbf{X} & \textbf{High}      & \textbf{Low}     & \textbf{-}      & \textbf{Bound Checking, Randomized Memory Layout}     \\ \cline{2-10} 
                            & \textbf{Website Spoofing} & \textbf{\checkmark} & \textbf{\checkmark} & \textbf{\checkmark} & \textbf{X} & \textbf{High}      & \textbf{Low}     & \textbf{-}      & \textbf{Secure Websites}     \\ \cline{2-10} 
                            & \textbf{Malicious Code Injection} & \textbf{\checkmark} & \textbf{\checkmark} & \textbf{X} & \textbf{X} & \textbf{High}      & \textbf{Low}     & \textbf{-}      & \textbf{Proper Secure Coding}     \\ \cline{2-10} 
                            & \textbf{SQL Injection} & \textbf{\checkmark} & \textbf{\checkmark} & \textbf{X} & \textbf{X} & \textbf{High}      & \textbf{Low}     & \textbf{-}      & \textbf{Proper Secure Coding}     \\ \cline{2-10} 
                            & \textbf{Cross-Site Scripting} & \textbf{\checkmark} & \textbf{\checkmark} & \textbf{\checkmark} & \textbf{X} & \textbf{High}      & \textbf{Low}     & \textbf{-}      & \textbf{Proper Secure Coding}     \\ \cline{2-10} 
                            & \textbf{Cross-Site Request Forgery} & \textbf{\checkmark} & \textbf{\checkmark} & \textbf{X} & \textbf{X} & \textbf{High}      & \textbf{Low}     & \textbf{-}      & \textbf{Proper Secure Coding}     \\ \hline
\multirow{7}{*}{\textbf{Application Attacks}}  & \textbf{Scareware} & \textbf{\checkmark} & \textbf{\checkmark } & \textbf{\checkmark } & \textbf{\checkmark } & \textbf{High}      & \textbf{Low}     & \textbf{-}      & \textbf{Anti-Malware, Anti-Scareware/Ransomware}     \\ \cline{2-10} 
                            & \textbf{Spyware} & \textbf{\checkmark} & \textbf{X} & \textbf{X} & \textbf{X} & \textbf{High}      & \textbf{Low}     & \textbf{-}      & \textbf{Anti-Spyware}     \\ \cline{2-10} 
                            & \textbf{Trojan} & \textbf{\checkmark} & \textbf{\checkmark} & \textbf{\checkmark} & \textbf{X} & \textbf{High}      & \textbf{Low}     & \textbf{-}      & \textbf{Anti-Virus}     \\ \cline{2-10} 
                            & \textbf{Botnet} & \textbf{\checkmark} & \textbf{X} & \textbf{\checkmark} & \textbf{X} & \textbf{High}      & \textbf{Low}     & \textbf{-}      & \textbf{Anti-Malware Software}     \\ \cline{2-10} 
                            & \textbf{Virus} & \textbf{\checkmark} & \textbf{\checkmark} & \textbf{\checkmark} & \textbf{X} & \textbf{High}      & \textbf{Low}     & \textbf{-}      & \textbf{Anti-Virus}     \\ \cline{2-10} 
                            & \textbf{File Infection} & \textbf{\checkmark} & \textbf{\checkmark} & \textbf{X} & \textbf{X} & \textbf{High}      & \textbf{Low}     & \textbf{-}      & \textbf{Anti-Virus, Anti-Malware}     \\ \cline{2-10} 
                            & \textbf{Rootkit} & \textbf{\checkmark} & \textbf{\checkmark} & \textbf{X} & \textbf{\checkmark} & \textbf{High}      & \textbf{Low}     & \textbf{-}      & \textbf{Anti-Malware, Stronger Authentication}     \\ \hline
\multirow{10}{*}{\textbf{Network Attacks}} & \textbf{Eavesdropping} & \textbf{\checkmark} & \textbf{X} & \textbf{X} & \textbf{X} & \textbf{High}      & \textbf{Low}     & \textbf{Lightweight Cryptography}      & \textbf{Firewalls, Proxies, VPN}     \\ \cline{2-10} 
                            & \textbf{Replay} & \textbf{X} & \textbf{\checkmark} & \textbf{\checkmark} & \textbf{X} & \textbf{High}      & \textbf{Low}     & \textbf{Lightweight Encryption, Hashing}      & \textbf{Timestamps, Attentive Filtering}     \\ \cline{2-10} 
                            & \textbf{Man-in-the-Middle} & \textbf{\checkmark} & \textbf{\checkmark} & \textbf{X} & \textbf{X} & \textbf{High}      & \textbf{Low}     & \textbf{Lightweight Encryption, Hashing}      & \textbf{Certificate Authority}     \\ \cline{2-10} 
                            & \textbf{Packet Sniffing} & \textbf{\checkmark} & \textbf{X} & \textbf{\checkmark} & \textbf{X} & \textbf{High}      & \textbf{Low}     & \textbf{Real-Time Packet/Data Encryption}      & \textbf{IDS}     \\ \cline{2-10} 
                            & \textbf{Password Cracking} & \textbf{\checkmark} & \textbf{X} & \textbf{X} & \textbf{\checkmark} & \textbf{High}      & \textbf{Low}     & \textbf{Strong Encryption Mechanisms}      & \textbf{Strong Multi-Factor}     \\ \cline{2-10} 
                            & \textbf{Traffic Analysis} & \textbf{\checkmark} & \textbf{X} & \textbf{X} & \textbf{X} & \textbf{High}      & \textbf{Low}     & \textbf{Lightweight Real-Time Encryption}      & \textbf{Secure Communication Channels, IDS, Network Filters}     \\ \cline{2-10} 
                            & \textbf{Wireless Jamming} & \textbf{X} & \textbf{X} & \textbf{\checkmark} & \textbf{X} & \textbf{High}      & \textbf{Moderate}     & \textbf{-}      & \textbf{IDS, Firewalls, Frequency Hopping/Shifting}     \\ \cline{2-10} 
                            & \textbf{Black-hole} & \textbf{X} & \textbf{X} & \textbf{\checkmark} & \textbf{X} & \textbf{High}      & \textbf{High/Moderate}     & \textbf{X}      & \textbf{Timer Based Baited Technique}     \\ \cline{2-10} 
                            & \textbf{Byzantine} & \textbf{X} & \textbf{X} & \textbf{\checkmark} & \textbf{\checkmark} & \textbf{High}      & \textbf{Moderate/Low}     & \textbf{-}      & \textbf{Firewalls, IDS/IPS, Honeypots}     \\ \cline{2-10} 
                            & \textbf{DoS/DDoS} & \textbf{X} & \textbf{X} & \textbf{\checkmark} & \textbf{X} & \textbf{High}      & \textbf{High/Moderate}     & \textbf{Lightweight Encryption}      & \textbf{Timestamps, IDS/IPS, Firewalls}     \\ \hline
\multirow{6}{*}{\textbf{Infrastructure Attacks}}  & \textbf{Insider Attack} & \textbf{\checkmark} & \textbf{\checkmark } & \textbf{\checkmark } & \textbf{\checkmark } & \textbf{Very High}      & \textbf{Moderate}     & \textbf{-}      & \textbf{Employee Screening, Constant Training, NDA}     \\ \cline{2-10} 
                           & \textbf{Outsider Attack} & \textbf{\checkmark} & \textbf{\checkmark} & \textbf{\checkmark} & \textbf{\checkmark} & \textbf{High}      & \textbf{Low}     & \textbf{-}      & \textbf{Closing Unwanted Ports, Traffic Monitoring}     \\ \cline{2-10} 
                            & \textbf{Key Stroke Register} & \textbf{\checkmark} & \textbf{X} & \textbf{X} & \textbf{X} & \textbf{High}      & \textbf{Moderate}     & \textbf{-}      & \textbf{BYOD Policy, Disable Any USB Port, Prevent Any Untrusted Software}     \\ \cline{2-10} 
                            & \textbf{USB Infection} & \textbf{\checkmark} & \textbf{\checkmark} & \textbf{\checkmark} & \textbf{\checkmark} & \textbf{Very High/High}      & \textbf{High/Moderate}     & \textbf{-}      & \textbf{BYOD Policy, Disable Any USB Port}     \\ \cline{2-10} 
                            & \textbf{Social Engineering} & \textbf{\checkmark} & \textbf{X} & \textbf{X} & \textbf{\checkmark} & \textbf{High}      & \textbf{Low}     & \textbf{-}      & \textbf{Employee Training, Awareness \& Accounting}     \\ \cline{2-10} 
                            & \textbf{Reverse Engineering} & \textbf{\checkmark} & \textbf{X} & \textbf{X} & \textbf{\checkmark} & \textbf{High}      & \textbf{Low}     & \textbf{-}      & \textbf{Employee Training, Awareness \& Accounting}     \\ \hline
\end{tabular}
\end{sidewaystable*}


\section{Ethical Hacking}
\label{sec:4}
In the era where ethical hacking is gaining a great fame and popularity, it is important to identify its life-cycle. Moreover, the tools used in its life-cycle events must also be highlighted on in order to know how the process proceeds. However, it is important to know the main drawbacks and challenges that surround the ethical hacking domain, first.

\subsection{Ethical Hacking Life-cycle}
\begin{figure*}[!ht]
  \centering
  \begin{minipage}[b]{0.75\textwidth}
    \includegraphics[width=\textwidth]{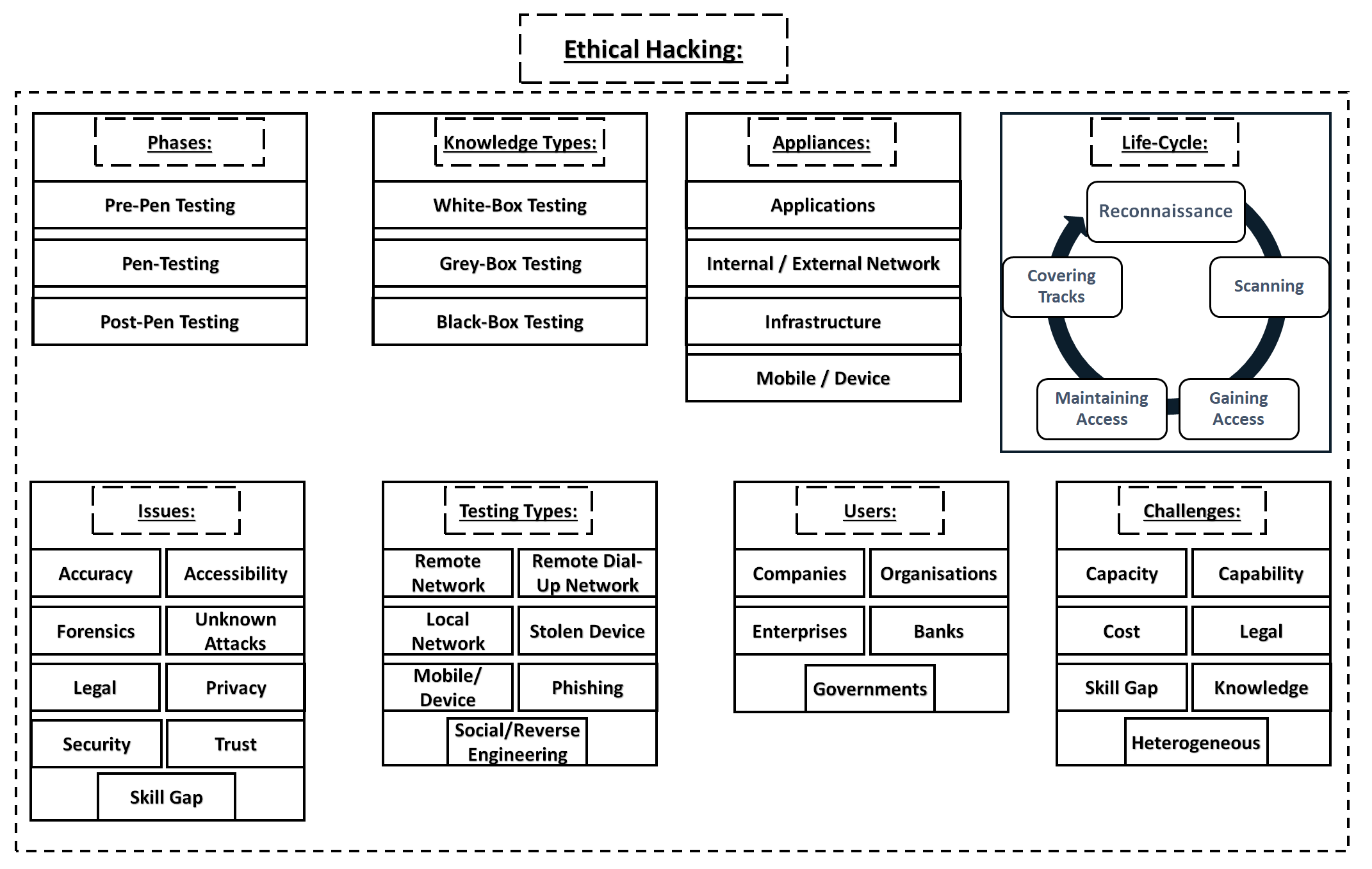}
    \caption{Ethical Hacking Classification}
    \label{fig:b}
  \end{minipage}
\end{figure*}

Ethical hacking life-cycle is closely similar to normal hacking life-cycle with only a slight difference based on the fact that ethical hackers aim to cause no harm nor damage to a given system, unlike black-hat hackers. However, the ethical hacking life-cycle (see \figurename~\ref{fig:b}) is divided into five main phases. This includes reconnaissance, scanning, gaining access, maintaining access, and covering tracks. In fact, in many cases, the ethical hacking cycle can further proceed into a non-ending loop as part of a constant risk assessment.
\begin{itemize}
\item \textbf{Reconnaissance:}
The reconnaissance phase is based on the use of the available processes and techniques that can either be used covertly, or deliberately in order to gather information about systems or/and users. In this phase, ethical hackers rely on passive attacks (i.e eavesdropping) to gather information about the network covertly over a long period of time. Moreover, it can be done actively by gathering as much information as possible, however, at the risk of being detected. Information gathering can either be done physically, where it can take some form of stalking to gather as much information as possible through social and reverse engineering, or logically, based on intercepting and sniffing network packets. The information gathering process aims to identify the  Wi-Fi in use, type of machines, along their software, hardware and Operating System (OS), along with their level of protection. Moreover, it also includes any open available port that can be used through network mapping.

\item \textbf{Scanning:}
is the next step that ethical hackers use and rely on for further vulnerability exploitation through simulated attacks. Scanning is based on conducting pen testing to discover any security or/and vulnerability gap(s) that can be used to conduct the attack. This includes the search for open or/and unused open ports, live hosts, devices, systems and services, along configuration/security vulnerabilities in firewalls, Intrusion Detection Systems/ Intrusion Preventing Systems (IDS/IPS), as well as routers and switches. Once a full picture is built over the system and the vulnerabilities are already identified. The second part of scanning, which is known as enumeration, takes place to gather information about a given target machine, device, system, or service. This is done by maintaining an active connection with it.

\item \textbf{Gaining Access:}
Once vulnerabilities are identified, with all the needed information being gathering ethical hackers will then try to gain access. This is achieved by relying on various pen-testing tools and techniques to virtually break into the system and bypass the security measures. Such phase is based on retrieving passwords through cracking attacks. In fact, the available tools can also be used for password cracking attacks. Thus, targeting the authorisation or authentication of a given system depending on the need.

\item \textbf{Maintaining Access:}
Once access into a given system is gained, the system’s resources can be exploited by seeking other common vulnerable devices. This includes spreading a worm into them and infecting them, or spreading malware or viruses. As a consequence, this can turn these devices into bots or “zombies”, or to implement a rootkit. This ensures a remote access by the elevation of privilege. Thus, gaining an administrative privilege at both OS and application levels. However, overcoming such attacks not only requires the implementation of Firewalls and IDS/IPS, but also Honeypots. In fact, the choice of Honeypots seems to be the most effective way to trap any attacker attempting to gain and maintain access over a given system. Moreover, for further reading, Maintaining Access tools are presented~\cite{patil2017ethical}.

\item \textbf{Covering Tracks:}
After gaining and maintaining a successful attack, the attacker will then try to cover his tracks. This is achieved by relying on the use of forensic and anti-forensic techniques and tools. Such a reliance helps eliminating any evidence that can reveal the attacker’s tracks and identify them. Hackers usually rely on covering their tracks by deleting log files, audit files and registry files that contain the failed log-in attempts and suspicious network behaviours. They also rely on the use of anti-forensic tools and techniques to hide any source of evidence. Thus, eliminating any source of evidence. In fact, attackers rely on a given system to launch further attacks without being detected. This is achieved by turning the intended device into a bot to launch D-DoS attacks. Therefore, the covering-tracks phase can be the start of the new cycle.
\end{itemize}

\subsection{Ethical Hacking Tools}
In order to achieve the intended ethical hacking cycle, different tools should be used so ethical hackers can perform and conduct their simulated attack. Furthermore, \tablename~\ref{tab:5} presents a summary of the most famous tools used for ethical hacking purposes.

\begin{table*}[!htp]
\centering
\small
\caption{{Ethical Hacking Life-cycle Tools}}
\label{tab:5}
\begin{tabular}{|p{3cm}|p{2cm}|p{11cm}|}
\hline
\textbf{Ethical Hacking Cycle}                  & \textbf{Tools} & \textbf{Description} \\ \hline
\multirow{4}{*}{\textbf{Reconnaissance}} & { IP lookup } & { Allows the identification of the IP in use with a geographical location } \\ \cline{2-3} 
                           & { MAC lookup } & { Allows the identification of the type of the device and the manufacturer } \\ \cline{2-3} 
                           & { BSSID } & { based on geographically locating the device in real time
} \\ \cline{2-3} 
                           & { Ping } & { Based over giving both the website name, IP, and the geographical location } \\ \hline
\multirow{5}{*}{\textbf{Scanning}} & { Angry IP scanner } & { scans for online IPs and available hosts within a small range, medium range and up to a broadcast range (single or multi IP scan) } \\ \cline{2-3} 
                           & { Nmap } & { Network Mapper used for network discovery along security auditing } \\ \cline{2-3} 
                           & { Netcraft } & { allows application testing, anti-phishing, and anti-fraud services } \\ \cline{2-3} 
                           & { Tracert } & { ensures a network analysis and diagnosis, identifies the track of the sent packet from an IP address to another } \\ \cline{2-3} 
                           & { Znmap } & { GUI Nmap version for network diagnosis } \\ \hline
\multirow{5}{*}{\textbf{Gaining Access}} & { Aircrack } & { used for 802.11a/b/g WEP and WPA cracking, used for brute force and dictionary password attacks alike } \\ \cline{2-3} 
                           & { Fluxion } & { combination of technical and social engineering techniques, lures a victim to type a Wi-Fi password to the attacker relying on keystrokes } \\ \cline{2-3} 
                           & { John The Ripper } & { password cracking penetration testing tool used for dictionary attacks } \\ \cline{2-3} 
                           & { Wireshark } & { captures real-time network data packets before being displayed in readable } \\ \cline{2-3} 
                           & { THC Hydra } & { login cracker, supports numerous attacking protocols, very fast and flexible, offers an easy way to gain a remote unauthorized access to the system } \\ \hline
\multirow{2}{*}{\textbf{Maintaining Access}} & { Beast } & { Remote Administration Tool or a "RAT" horse used to create backdoors } \\ \cline{2-3} 
                           & { Metasploit } & { cyber-security framework that provides vital information about known security vulnerabilities, ensures penetration testing exploitation strategies, methodologies and plans } \\ \hline
\textbf{Covering Tracks}                  & { OSForensics } & { forensic tool used to delete the log files, audit files and registry files beyond recovery } \\ \hline
\end{tabular}
\end{table*}

\subsection{Ethical Hacking Challenges}
\begin{figure*}[!ht]
  \centering
  \begin{minipage}[b]{0.9\textwidth}
    \includegraphics[width=\textwidth]{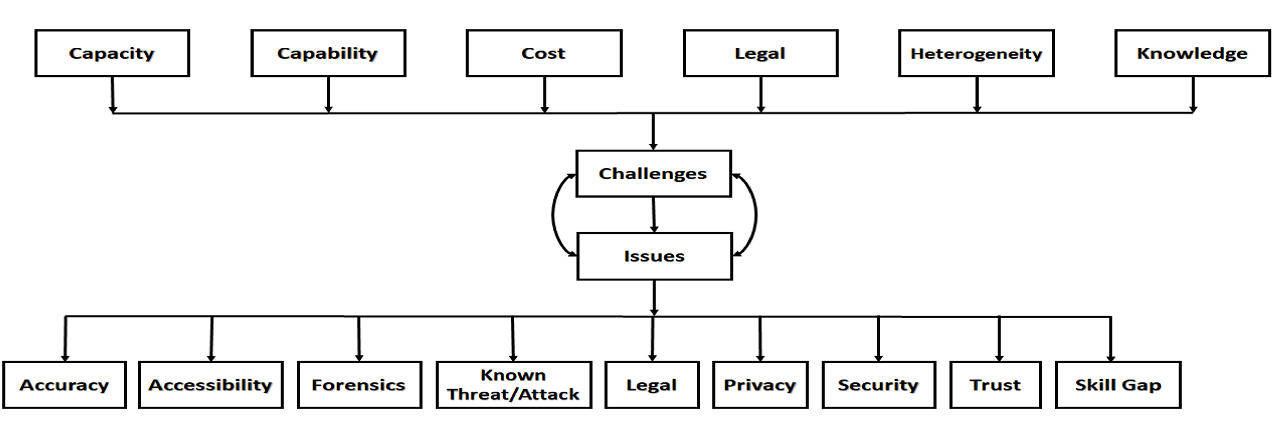}
    \caption{Ethical Hacking-Pen Testing Issues \& Challenges}
    \label{fig:c}
  \end{minipage}
\end{figure*}

There are various challenges that the surround the ethical hacking domain. In fact, it is very important to highlight them to identify the problem, verify the challenge and overcome them both. However, these challenges are not limited to one or two aspects, but to various different aspects instead.
Therefore, these challenges and issues that surround both, the ethical hacking and pen testing domains can be summarized in \figurename~\ref{fig:c}.

\begin{itemize}
\item \textbf{Capability Challenges:}
Many challenges are related to the lack of experience and skills gained over time. In fact, there’s a threshold difference between different ethical hacking teams~\cite{pike2013ethics}. Moreover, there’s no standardised threshold that unifies the skills and experience of ethical hackers into a single capability and capacity. Therefore, some ethical hacking teams may have more skills, experience, and knowledge, compared to other groups when performing their pen testing, along with the availability of a much more sophisticated tools and kits.

\item \textbf{Capacity Challenges:}
Another challenge is performing the necessary pen testing in order to evaluate the level of security and immunity of a given organisation against cyber-attacks, especially in terms of risk management~\cite{saleem2006ethical}. The capacity is based on the limited experienced manpower, and the available resources, used to perform the pen testing technique(s) and attack(s). Therefore, this is another challenge that requires a deeper focus and attention.

\item \textbf{Cost Challenges:}
The cost of performing a pen testing attack is not cheap. However, it is necessary to avoid any exploitation of any vulnerability or security gap~\cite{pike2013ethics}. In fact, pen testing is divided into two main steps. The first one requires the identification of already existing exploitable vulnerabilities which requires a defined cost. The next step is based on the ability to offer security measures to further protect the system, which also requires an additional cost. 

\item \textbf{Legal Challenges:}
Many legal challenges also surround the ethical hackers, along with the ethical hacking as well. In other terms, ethical hackers do not perform their pen testing without signing a legal document called the Non-Disclosure Agreement (NDA). This also requires notifying the required authorities so their testing is not classified as a cyber-crime. Therefore, without the signing of legal processes, ethical hackers risk being legally prosecuted and arrested~\cite{maurushat2019ethical}. 

\item \textbf{Heterogeneous Challenges:}
Such heterogeneous challenges are based on different ethical hacking groups performing different pen testing attacks and types from their perspectives and skills~\cite{maurushat2019ethical}. Though it is important and necessary, an exploitable vulnerability identified by one ethical hacking team, may not be identified by another ethical hacking team, and vice versa. Therefore, the choice of the right ethical hacking team to perform the right and necessary pen testing is somewhat of hard task and challenge.

\item \textbf{Knowledge Challenges:}
Knowledge challenges are based on the ability of ethical hackers to perform their pen testing against exploitable vulnerabilities and security gaps. This includes software bug, misconfiguration or other bugs (i.e hardware, configuration, or coding). However, their pen testing and knowledge are based on the ability to identify and overcome the only already existing attacks. In other terms, pen testing is unable to detect new attacks such a zero-day attacks~\cite{bilge2012before}. This is due to these attacks being based on exploiting a vulnerability that was not detected by ethical hackers who were conducting their pen testing. This also includes the birthday attack~\cite{coppersmith1985another,girault1988generalized} in a way, which is quite similar to zero-day. This is done, in addition to the presence polymorphic malware~\cite{you2010malware} and crypting services that keep on changing their signature and behaviour patterns. Thus, their identification and mitigation process is becoming seriously challenging. This is all due to their ability to avoid and evade being detected by intrusion detection systems, firewalls and anti-viruses.
\end{itemize}

\subsection{Ethical Hacking Issues}
Ethical hacking has been prone to lots of different issues that surround this domain, especially with ethical hackers lacking  skills, experience and knowledge for a start. There is also different varieties of performing different ethical hacking tests. As a result, these varieties have their own advantages and drawbacks. However, there are far more important issues that require attention, first.
\begin{itemize}

\item \textbf{Accuracy:}
the accuracy of the performed tests may not provide the right results, which in turn can be inaccurate. In some cases, the accuracy of the performed pen testing is somewhat of an exaggeration, claiming that a given system is 100\% secure, where in reality it is not, since no system is fully secure. In fact, the accuracy depends on each pen testing being conducted on every aspect of a given organisation using different scenarios to ensure that the system is secure enough against any attack attempt(s). Therefore, the accuracy is an issue itself. 
\item \textbf{Accessibility:}
in order to perform pen testing, ethical hackers require the need to access the given organisation logically, as well as physically. Such accessibility is usually led by a simulated attack. In fact, it depends on the organisation’s access rights and privileges offered for ethical hackers in order to perform their tested attack. In many cases, ethical hackers are given the least privileges (black box) for two main reasons such as testing their skills, and due to the organisation not wanting to reveal some of its information (part of the organisation’s security policy).

\item \textbf{Forensics:}
Many ethical hackers do lack the ability to use and rely on using forensic tools and techniques in order to retrieve any data from logs and audits to check for any failed login attempt(s), or abnormal network behaviour. In some cases, ethical hackers rely on the use of anti-forensic techniques in order to simulate an advanced attack to cover the attacker’s track(s). This method is performed and applied in order to check the ability of a given organisation to defend, detect and locate the attacker. In fact, it is classified as another issue, where most ethical hackers have a lack of knowledge in the digital forensic domain especially in web-applications, computer, mobile and cloud forensics~\cite{ruan2012cybercrime,de2016pentesting,farsole2010ethical}.

\item \textbf{Known Attacks:}
another limitation that surrounds the ethical hacking and pen testing domains, is the limited ability of ethical hackers to conduct pen testing. This is done in order to evaluate a given organisation's security measures against already known attacks. However, it cannot predict unknown attacks such as zero-day and polymorphic malware. This is another serious limitation that does eventually affect the organisation's security, especially since the main threat and risk come from an unknown attack.

\item \textbf{Legal:}
issues are highly emerging, especially since ethical hackers need to perform a simulated cyber-attack or/and hacking which, in real cases, would be classified as a cyber-crime~\cite{harper2011gray}. Therefore, the reliance on signing a non-disclosure agreement to protect both sides is mandatory. Moreover, when performing simulated insider attacks, police law enforcement and local authorities must be notified, so that in case the organisation managed to capture an ethical hacker, there will not be any prosecution(s) nor arrest(s). Therefore, this is another important issue that requires some serious attention. 

\item \textbf{Privacy:}
The privacy issue must be taken very seriously. This is due to the fact that the organisation, which is relying on ethical hackers to perform their pen testing, is under a simulated attack. Such a simulated attack mainly targets the privacy of this organisation including data, information, and system’s privacy~\cite{jamil2011ethical}. In fact, this also includes targeting their confidentiality, integrity and availability. Such simulated attacks are based on a privacy breach, which can be serious if it was not conducted by ethical hackers. Therefore, ethical hackers need to explain to the organisation each step being performed, in respect to their privacy. Otherwise, it will remain as constant issue.

\item \textbf{Security:}
Security issues are more related to the lack of experience, knowledge and skills among the ethical hacking team that is performing the pen testing~\cite{brey2007ethical}. It is also related to the weak security measures and access control mechanisms employed by a given organisation to protect itself from any given physical access attempt(s). Moreover, many ethical hacking teams’ lack or does not offer any security suggestions nor recommendations. Thus, relying on the commercial part of conducting ethical hacking for commercial and financial purposes.

\item \textbf{Trust:} issues are primary related to fully trusting a third (suspicious) party~\cite{thomas2017issues} in order to evaluate a company’s security level(s)~\cite{sahare2014study}. In fact, performing pen testing would allow ethical hackers to know all information about the organisation. This includes their exploitable vulnerabilities and available security gaps. The main concern is finding among the ethical hackers’ team a black-hat or a grey-hat hacker capable of leaking these sensitive information to malicious third party (rival organisations(s), or cyber-criminal groups) without the company’s knowledge. This would lead to a far more greater damage, implications, serious consequences, and issues alike.

\item \textbf{Skill Gap:}
Another issue have risen, especially with the highly increasing number of hackers. This issue is related to skill gap issues~\cite{rafferty2016dangerous}. In fact, both cyber-criminals and hackers rely on it in order to perform their cyber-attacks more freely and more easily. This can be conducted without encountering any serious defensive security measures employed based on the recommendations of ethical hackers who are assessing the level of security of a given organisation. 
\end{itemize}


\section{Pen Testing}
\label{sec:5}
Pen testing require the presence of ethical hackers to perform this task in a verified professional way that guarantees their safety and the organisation's safety. Therefore, it is highly important to know how pen testing is linked to ethical hackers, and how ethical hackers perform their pen testing duties and tasks.\\

\begin{figure*}[!ht]
  \centering
  \begin{minipage}[b]{0.9\textwidth}
    \includegraphics[width=\textwidth]{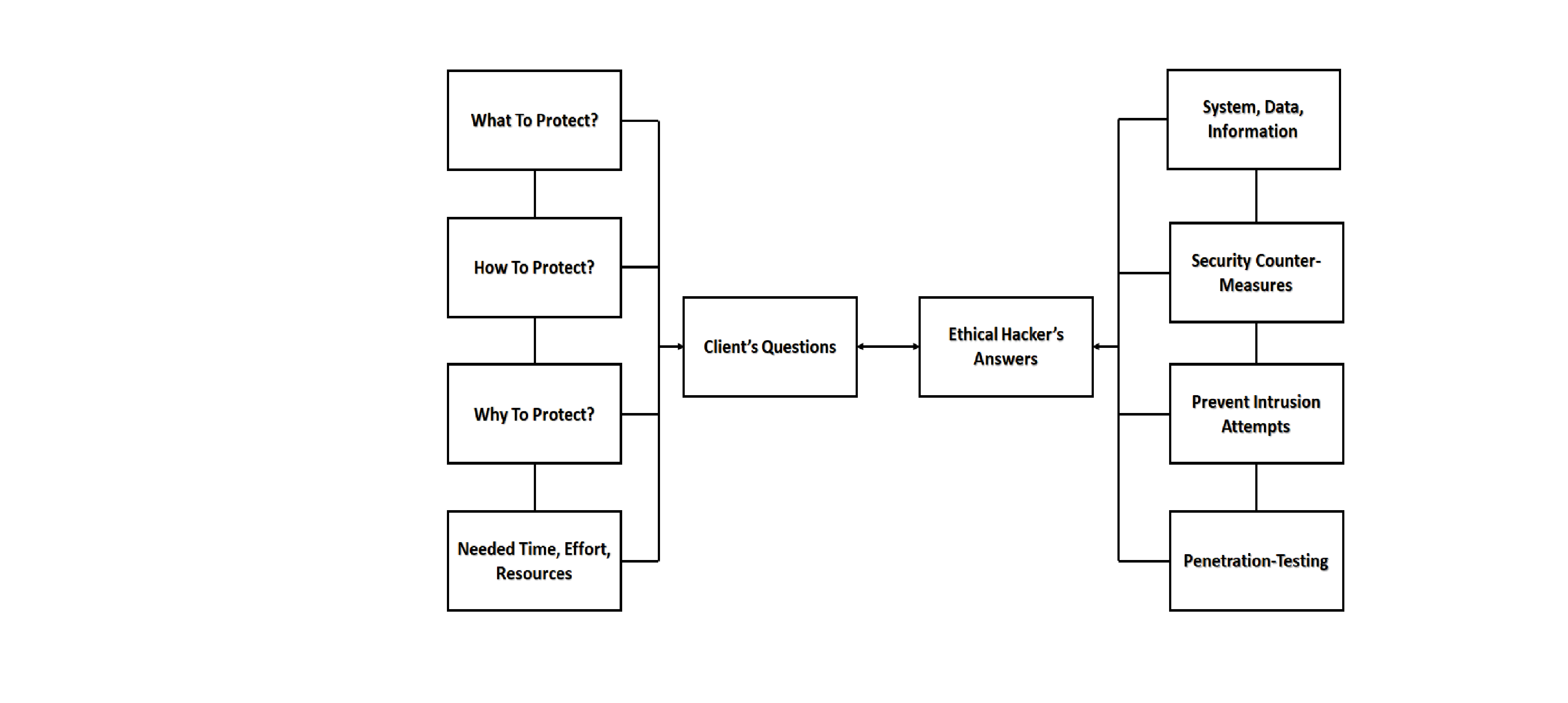}
    \caption{Answering The Questions}
    \label{fig:d}
  \end{minipage}
\end{figure*}

Penetration tests, or pen-tests are being conducted with the purpose of evaluating the levels of security and immunity of a given IT and non-IT systems and personnel against any security gap or exploitable vulnerability~\cite{lunne2014cone} from an already known and possibly unknown attack. Thus, answering their customers needs (see \figurename~\ref{fig:d}). In fact, this vulnerability can possibly exist in an OS due to design flaw, mis-configuration, software bug or even failure. Therefore, the aim of pen testing is to safeguard the information from any alteration, modification or disclosure from already known threats and/or attacks. To fulfill this task, different security measures were suggested and presented by different papers including in~\cite{mcdermott2001attack,chen2017guarding}. Such papers, presented the implementation of the right Access Control mechanisms, along with a strong authentication process. This was accompanied by the implementation of additional firewalls, Intrusion Detection/Intrusion Prevention systems, along cryptographic mechanisms to protect data’s Confidentiality, Integrity and Availability (CIA).
\subsection{Pen Testing Pros \& Cons}
Pen testing has its own advantages and drawbacks that one must know and understand before diving deeper into the pen testing domain.
\begin{itemize}
\item \textbf{Pen Testing Pros:}
The main advantages of conducting pen testing, is based on ensuring the right evaluation through testing to achieve and fulfil a proactive security approach~\cite{saleem2006ethical,lunne2014cone,geer2002penetration}. Such approach allows the exploration and detection of security risks accurately, whilst ensuring a system’s adaptation to real-time changes. Moreover, pen testing also helps investigating data breaches and network intrusions. In fact, pen testing can find already known and in some cases, unknown hardware/software design bugs, flows, and misconfigurations that can be exploited using hacking tools. Therefore, assessing a given vulnerability, whilst evaluating both likelihood and impact of a given risk from occurring is a top priority and a must. Furthermore, pen testing ensures an information gathering process before testing a given system. Thus, learning all the essentially and needed information before conducting the given test. Such a pen test also allows testing the readiness of IT staff against a given attack, along with the level of training that the employees did undergo. This helps monitoring their ability and capability to resist against social engineering and phishing attacks. In fact, these simulated attacks are conducted not to cause harm, but to evaluate the degree of readiness, whilst also highlighting on the depicted security gap(s).
\item \textbf{Advanced Pen Testing:}
Many previous pen testing limitations were over-passed. As a result, advanced pen testing was introduced in~\cite{allen2012advanced} through the reliance on the latest available security exploitation and research. This serves as a significant advantage. Unlike traditional pen testing, the organisation is made aware of the importance and reliance on solid, well-built incident response team and program to adhere to any emerging threat before it even occur. In fact, it is all due to realising that security breaches are always there. Therefore, different training sessions, workshops, projects and conferences are being maintained to further educate ethical hackers to further investigate any possible intrusion~\cite{mowbray2013cybersecurity}.

\item \textbf{Pen Testing Cons:}
Despite highlighting on the advantages related to pen testing in~\cite{lunne2014cone,saleem2006ethical,geer2002penetration}, there are also disadvantages that require mentioning. This includes the limitation of time, where each test requires a specific amount of time to be performed, especially with no fixed time to finish conducting a given test when dealing with large organisations. In fact, such a process can range from days, and sometimes it can take up to weeks, months, and even years. Another drawback is based on the limitation of scope, methods and skills. The scope is limited to the resource, budget and security constraints. This prevents ethical hackers from performing all the intended and required testing. Moreover, the methods in use for pen testing can lead the system to crash. This may cause damage and harm to the system, which is not the aim and intent of conducting a pen testing. As for the skills, professional pen testers are limited in terms of certified/licensed members, available skills and manpower, despite that skilled pen testers rely on using a specific technology in their expertise,  field and domain. 
In fact, this is also a problem that digital forensic investigators also suffer from. In addition, ethical hackers have a restricted access right to a given targeted environment, due to a given organisation’s policy wanting to conduct external remote network pen testing. This problem also falls into the domain of experienced ethical hackers to cover all possible aspects in pen testing domain. Furthermore, the major drawback of pen testing is based on the pen testers’ ability and capability to conduct their pen testing against only already known attacks and threats. Thus, rendering them unable and useless when it comes to dealing with unknown threats and attacks. Finally, the last drawback of pen testing is having a uniform thinking over conducting a given pen testing without thinking outside the box. This prevents them from finding new methods and mechanisms to use. As a result, their tasks are limited, unlike attackers who rely on creating new methods to evade detection and perform their cyber attacks. Moreover, \tablename~\ref{tab:6} summarises the main pros and cons of pen testing.

\end{itemize}

\begin{table*}[!ht]
\centering
\small
\caption{{Pen Testing Pros \& Cons }}
\label{tab:6}
\begin{tabular}{|l|l|}
\hline
\multicolumn{2}{|c|}{\textbf{Pen Testing}} \\ \hline
\textbf{Pros}      & \textbf{Cons}      \\ \hline
{Accurate Detection Of Security Risks}      & {Time Consuming}      \\ \hline
{Help Investigation Network Intrusions/Breaches}      & {Useful Against Already Known Attack/Threats}      \\ \hline
{Help Assessing Vulnerabilities}      & {Lack Of Innovative Methods}      \\ \hline
{Very Useful Against Known Attacks}      & {Limited Access Rights}      \\ \hline
{Detect Security Gaps}      & {Limited Skills}      \\ \hline
{Evaluate The Organisation’s Security Level}      & {Already Known Methods In USe}      \\ \hline
{Evaluate the Response Level Of IT \& Employee Staff}      & {Budget Security Constraints}      \\ \hline
{Simulated Evaluative Attacks}      & {Possibility of Causing System’s Harm}      \\ \hline
\end{tabular}
\end{table*}

\subsection{Pen Testing Phases}
Pen testing is eventually divided into three main phases. This includes the pre-pen testing, pen testing, and post-pen testing phases. In fact, each phase is dedicated to perform its required work. This is done in collaboration and cooperation of other phases to ensure a successful pen-testing.


\begin{itemize}

\item \textbf{Pre-Pen Testing:}
is based on ensuring a meeting between the ethical hacker and the organisation’s representative(s) to discuss and agree over the conducted ethical hacking tests. This includes a simulated non-damaging attack against the intended organisation. After discussing and agreeing over the specific tasks, objectives and goals, both sides sign a non-disclosure agreement to protect ethical hackers when conducting their simulated pen-test attack in a legal manner.

\item \textbf{Pen Testing:}
after setting up the objectives and goals, ethical hackers will divide themselves into two teams. The first team is the “blue team” who intends to defend a given system. The second team is the “red team” that intends to break into the system, by trying to overcome the IT security staff and countermeasures. Once teams are divided, the pen-testing attack takes place following the already ethical hacking cycle.

\item \textbf{Post-Pen Testing:}
once the pen-testing has been conducted and finished, a security level evaluation based on vulnerability assessment will be conducted. Afterwards a physical report (rather than logical report for security purposes) will be handed to the organisation’s representative(s). The report will reveal the tools and attacks used and conducted, along with the assessment of the security level of a given organisation. This report also includes the security gaps and vulnerabilities found, detected and exploited. Moreover, it also includes suggestions and recommendations that will be handed over to an organisation to increase its security measures.
\end{itemize}

\subsection{Pen Testing Tools}
Different Tools are being used in order to ensure the most effective and successful pen-testing. In fact, different papers~\cite{de2016pentesting,de2016pentesting} presented the most known tools that are mainly used, and which will be highlighted in \tablename~\ref{tab:7}, which  is a summary of the most used pen testing tools, aside from mentioning in earlier tables that different tools were used for VAPT. In fact, a brief comparison between vulnerability assessment and penetration testing was made in~\cite{tang2014guide}.
\begin{table*}[!ht]
\centering
\small
\caption{\textbf{Famous Pen Testing Tools }}
\label{tab:7}
\begin{tabular}{|p{4cm}|p{11cm}|}
\hline
\textbf{Pen Testing Tools} & \textbf{Description} \\ \hline
\textbf{Acunetix} & Scans and detects web vulnerabilities including XSS , SQL Injection, Code execution and File inclusion/upload form \\ \hline
 \textbf{Burp Suite} & Used for Pen-Test and security websites, capable of Proxy Interception, write plugins, and Automatic vulnerability detection \\ \hline
\textbf{Backbox Linux} & focuses on pen testing and safety assessment and penetration testing, mainly used for the analysis of web applications and networks  \\ \hline
Kali Linux & An advanced Pen Testing and Security Auditing, with hundreds of tools to perform PenTesting, Forensics and Reverse Engineering, and other wireless, web an network security analysis \\ \hline
\textbf{Nessus} & Deals with a large number of network devices, also used for web applications Pen testing \\ \hline
Parrot Security OS & Highly efficient lightweight OS designed for ethical hacking, computer forensics testing, cryptography and other security tasks \\ \hline
\textbf{Samurai Web Testing Framework} & Preconfigured framework that functions as a web pen -testing platform, capable of detection website code vulnerabilities\\ \hline
\textbf{SQLmap} & Automated command line capable of detecting and exploiting any SQL Injection vulnerability, whilst extracting information from databases\\ \hline
\end{tabular}
\end{table*}

\subsection{Pen Testing Knowledge Types}
\begin{figure*}[!ht]
  \centering
  \begin{minipage}[b]{0.7\textwidth}
    \includegraphics[width=\textwidth]{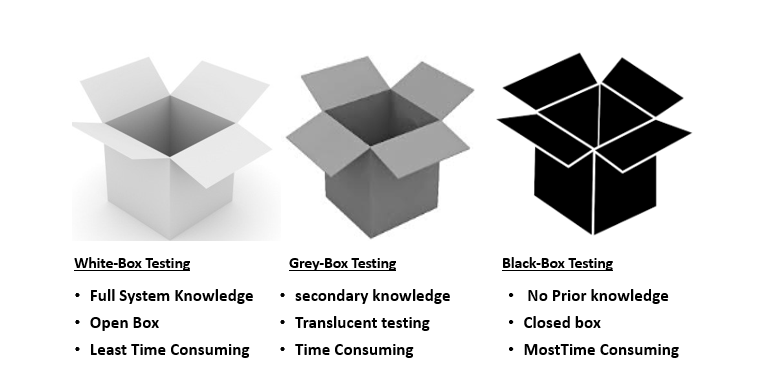}
    \caption{Box-Testing}
    \label{fig:e}
  \end{minipage}
\end{figure*}

Pen-testing knowledge is usually divided into three main types~\cite{caldwell2011ethical,turpe2009testing,al2018study}. Each type is classified as part of the conducted pen-testing in order to evaluate the level of security and immunity of a given organisation in each tested case. These three types are known as white box testing, grey box testing, and black box testing (see \figurename~\ref{fig:e}). Such types are based on the hacker’s level of knowledge against its target. It can be based on having full knowledge (white box), partial knowledge (grey box), or no knowledge (black box) of the system at all. Hence the reliance on penetration testing to evaluate the system against such attack types.

Moreover, box testing technique are presented in \tablename~\ref{tab:a}.

\begin{table*}[!ht]
\centering
\small
\caption{{Box Testing Techniques~\cite{nidhra2012black,khan2012comparative}}}
\label{tab:a}
\begin{tabular}{|p{3cm}|p{4cm}|p{9cm}|}
\hline
\textbf{ Testing Type }                  & \textbf{ Testing Technique } & \textbf{ Description } \\ \hline
\multirow{5}{*}{\textbf{ White Box Testing }} & { Statement Coverage } & {Used to test every statement more than once. Tool in use: Cantata++ } \\ \cline{2-3} 
                           & { Decision Coverage } & { Used to test every decision condition including conditional loops more than once. Tool in use: TCAT-PATH } \\ \cline{2-3} 
                           & { Condition Coverage } & { Ensures a one-time code execution mandatory once conditions are tested } \\ \cline{2-3} 
                           & { Decision/Condition Coverage } & { Used to test all the Decision/Condition coverage during the code execution } \\ \cline{2-3} 
                           & { Multiple Condition Coverage } & { Each system entry point is executed more than once } \\ \hline
\multirow{5}{*}{\textbf{ Black Box Testing }} & { Finds errors in the input values boundaries} & { Finds errors in the input values boundaries} \\ \cline{2-3} 
                           & { Equivalence Class Partitioning} & { Excessive application testing to input redundancy. Inputs divided into classes with values for each class} \\ \cline{2-3} 
                           & { Decision Table Based Testing} & { Employed when the action is applied under unstable varying conditions} \\ \cline{2-3} 
                           & { Cause-Effect Graphing Technique} & { Operates on a system’s external behaviour only, and helps selecting and creating test cases} \\ \cline{2-3} 
                           & { Error Guessing} & { Success rate depends on the experience of the tester. Test cases are written while reading a document or when encountering an undocumented error} \\ \hline
\multirow{4}{*}{\textbf{ Grey box Testing }} & { Matrix Testing  } & { Defines all the existing variables on a given program } \\ \cline{2-3} 
                           & { Regression Testing } & { Relies on retest all, retest risky use cases, retest within a firewall strategies to check if previous changes regressed other program’s new version aspects } \\ \cline{2-3} 
                           & { Pattern Testing } & { Goes through the code to identify the cause of failure by retrieving historical data of the previously defected system } \\ \cline{2-3} 
                           & { Orthogonal Array Testing } & { Offers the most code coverage with the least test cases } \\ \hline

\end{tabular}
\end{table*}

\begin{itemize}
\item \textbf{White-Box Testing:}
is also known as open box testing~\cite{nidhra2012black}. It is based on having full knowledge regarding the intended and targeted system in addition to all its software and firmware components. White box testing is based on the attacker’s ability to have an insider with a full knowledge over how a program runs. However, White-Box testing is less time consuming since there's a prior full knowledge of the system. This makes it very suitable for testing algorithms. It also ensures that all logical decisions are being verified, and ensures syntax checking with any possible design error.


\item \textbf{Grey-Box Testing:}
 is also known as translucent testing. It is based on having a secondary knowledge about the system. In other words, having a partial knowledge over a given system. In grey-box testing, the internal programming is partially known. That’s the reason why the secondary knowledge is maintained. Unlike White box testing, it is not time consuming~\cite{khan2012comparative}. However, it is not suitable for testing algorithms. In fact, it is non-intrusive and unbiased, with the least risk of conflicts between testers and developers.

\item \textbf{Black-Box Testing:}
can also include application black-box testing~\cite{bau2010state}. In fact, is based on the fact that testers do not have any prior knowledge regarding the internal/external system, nor over the software and firmware structure used~\cite{nidhra2012black}. Moreover it can also be tested on android applications~\cite{zhauniarovich2015towards}. As a result, such a test is conducted based on the user’s perspective and not the designer’s perspective. Despite that the program used is not known. A pen-tester may be an expert, since testers rely on any system contradictions. This testing form requires the most time consumption as there is no prior knowledge to the tested system. However, it is not useful for testing algorithms,nor complex segments of code since many program paths are left untested. In fact, it is also known as \textbf{Behavioural Testing} and can either be functional or non-functional.
\end{itemize}
\subsection{Pen Testing Types}
Pen testing has different types depending on ethical hackers who apply different testing types~\cite{palmer2001ethical}. However, a slight modification was made, where any of their combination testing may be required to perform the following tasks. This includes a remote network and dial-up network testing, local network testing, stolen device testing, social/reverse engineering testing, physical entries testing, along phishing (spear-phishing, vishing and whaling) testings, along others which will be presented as follows. That is why, pen testing is highly important and recommended~\cite{bishop2007penetration}.

\begin{itemize}
\item \textbf{Network Penetration Testing:}
provides security and protection for the entire network and network components against potential future attacks by identifying vulnerabilities found within the network.


\item \textbf{Mobile Application Penetration Testing:}
must satisfy both security and privacy requirements since Mobile applications are a critical business component which require from customers to trust mobile applications to have access and preserve their sensitive and personal information.


\item \textbf{Remote Network Testing:}
is based on simulating an intruder that is launching a cyber-attack. The aim of the intruder is to overcome firewalls, filtering routers and web servers to ensure a successful attack. Hence, this testing phase is based on overcoming the initial lines of defence employed by a given organisation. This helps in testing and checking their ability to overcome such intrusion attacks and attempts.

\item \textbf{Remote Dial-Up Network Testing:}
Such remote dial-up network testing is coordinated with local telephony companies. It also aims at simulating the intrusion attempt by an intruder. Such intruder aims at leading a cyber-attack against a given client’s modem pool. The intruder's aim is to defeat user authentication schemes. This type of remote testing is based on testing the effectiveness level of the employed authentication schemes to overcome similar attack types.

\item \textbf{Local Network Testing:}
The goal of Local Network Testing is to simulate a rogue employee (whistle-blower) who already has an authorised access to a given organisation’s network in order to lead an insider attack. Such a rogue employee must be able to overcome intranet firewalls, e-mail systems, server security measures, and internal web servers. However, local networks can be done through the employment of a rogue USB (mainly rubber ducky) that injects a malware into the system to record keystrokes, or/and implement a Rootkit or a Remote Access Trojan (RAT)~\cite{thompson2005spyware,chiang2007case}. Hence, it is very important to focus on this specific insider threat to avoid serious effects, and re-enforce a strong Bring Your Own Device (BYOD) policy.

\item \textbf{Stolen Device Testing:}
This type of testing is based on stealing a key employee’s device, including administrators, managers, or even coordinators. Such testing is based on stealing the device by a hacker masqueraded as a client without the employee’s knowledge. Once the device is stolen, it will be handed to the ethical hackers who will act as black-hat hackers. Once handed over, the device is scanned for any stored passwords, corporate information and personal information, in addition to company's sensitive information. Therefore, allowing hackers to gain a corporate intranet with the owner’s full access privilege.  

\item \textbf{Social \& Reverse Engineering Testing:}
Social engineering is based on targeting humans rather than machines, through the  exploitation of human nature and emotion~\cite{workman2008wisecrackers,krombholz2015advanced} . More precisely, it can be based on baiting and pretexting. Baiting is the ability to send gifts as gestures for a given employee to exploit their emotional feelings and urge their to reveal business secrets and information without their realisation. Pretexing is based on phoning the IT staff claiming to be a colleague, or a close cyber-security company that wants to cooperate with them in order to retrieve information as much as possible without their knowledge or realisation. This can also be called vishing. Reverse engineering is a different aspect, where the hacker relies on impersonating a colleague and starts asking lots of questions as a part of cooperating and collaborating in order to gain all the needed and essential information through information gathering~\cite{irani2011reverse}. In fact, both techniques can be used by a given hacker who acts as a lost employee in a given company, or an employee who forgot his ID card. However, no security measures are recommended except to raise the level of awareness and training. 

\item \textbf{Physical Entry Testing:}
Physical entry testing is based on gaining a physical access to a given organisation, mainly the Head-Quarters (HQ). Additional arrangements must be made to avoid ethical hackers being arrested whilst performing their testing in case they failed to overcome the physical entry testing. This testing is having the ethical hackers evade and avoid detection by  masquerading  an employeed staff. Such a phase requires a pre-phase planning that allows hackers to watch where the company dumps its waste, or even befriend another employee to gain additional information. It also includes eavesdropping on the employee’s conversations in a given restaurant or café. The best defence is to implement the necessary security policies based on a well-trained and well-aware security guards, alongside privileged access control mechanisms, with additional Close Circuit TeleVision (CCTV) constant coverage and monitoring. Moreover, it also requires training employees on eliminating and completely destroying any possible waste that serves as an advantage for a hacker. In fact, well-defended systems should prevent any intruder type (including semi-intruders and semi-outsiders). In addition, a well-defended system should only allow authorised users to access the required information. Furthermore, the evaluation of a client’s system(s) is based on several phases described by Boulanger in~\cite{boulanger1998catapults}.
\item \textbf{Phishing \& Spear-Phishing Testing:}
 this kind of pen testing is not being highlighted properly. Therefore, it is important to mention it in order to raise the level of awareness and training among different cyber-security and ethical hacking peers, researchers and colleagues alike. More focus must be aimed at social engineering and phishing attacks types~\cite{jagatic2007social,parmar2012protecting,caputo2014going}. Such attacks are based on hackers sending infected e-mails that include a fake company inviting them to cooperate and read more information about this organisation through the design of a fake website that lures the employee to click on it and visit it. Once visited, or once the "X" button is clicked on, a malware will be automatically installed into the employee’s device. Unlike phishing, spear-phishing is based on sending an infected fake Curriculum Vitae (CV) file in the form of Word or even PDF to IT staff. Meanwhile, whaling is based on mainly targeting administrators and Chief Executive Officers (CEOs). The best level of security measure is to raise the level of awareness and security among employees by avoiding malicious emails and links from unknown sources. This also includes the reliance on different machine learning algorithms that are capable of learning and detecting similar covert masqueraded attacks. 
 \end{itemize} 

\subsection{Pen Testing Appliance}
Pen-testing can also be applied in other domains. This includes application and network domains. In fact, it is applied to evaluate the levels of security in a given application and network both internally and externally, and  avoid any possible exploitation. In other terms, pen testing can be applied in Application Pen Test, Internal \& External Network Pen Testing, and Infrastructure \& Application Pen Testing.
\begin{itemize}
\item \textbf{Application Pen Testing:}
is used in order to reveal any possible application weaknesses and flaws, through the reliance on real-case strategies and attacks~\cite{singh2017penetration}. It is due to the fact there’s a close relation of cooperation and collaboration in-between applications and IT infrastructure through servers, networks and devices alike. Therefore, the appliance of such a pen-testing is based on identifying any possible exploitable vulnerability or/and security gap. It is achieved through the assessment and evaluation of both likelihood and impact of a given risk. Therefore, aiming to mitigate any given risk through a presented list of security recommendation(s) to fix a given issue and reduce the risk’s exposure.

\item \textbf{Internal \& External Network Pen Test:}
is usually conducted in order to reveal any existing network vulnerability by being able to perform a real-case attack scenarios~\cite{krutz2010cloud}. For example, a given scenario is based on the reliance on information gathering phases (either passively or actively, and either logically or passively). This also includes scanning for the right vulnerability to exploit with no intentions of causing any possible damage nor harm to the given intended system. Therefore, making the scenario as realistic as possible. Such a simulated attack would allow a given organisation to identify the path of the attack, as well as to identify and assess vulnerabilities according to the likelihood or/and impact of a possible risk that is to occur. Such a move is done through a risk mitigation. Thus, presenting the necessary recommendations to protect a network both internally and externally is advised.

\item \textbf{Infrastructure \& Application Pen Testing:}
is being performed on the cloud infrastructure domains (\figurename~\ref{fig:f}). Such  pen testing, includes both corporate applications and integrated networks~\cite{subashini2011survey}. Therefore, the aim is to help organisations identify any possible attack path(s), based on the identified security gap and exploitable vulnerability. Moreover, it is also able to assess both the likelihood and impact of a risk from occurring, before mitigating it. Furthermore, it offers the necessary recommendation(s) to enhance the level of security of a given organisation.
   \begin{figure*}[!ht]
  \centering
    \includegraphics[scale=0.5]{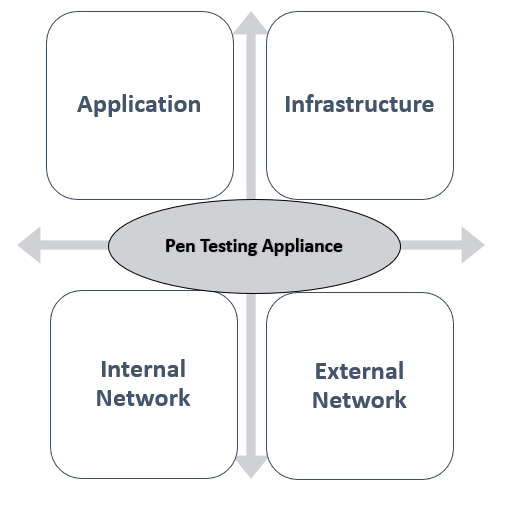}
    \caption{Pen Testing Appliance}
    \label{fig:f}
\end{figure*}
\end{itemize}


\subsection{Pen Testing Existing Solutions}
To ensure a much more accurate and successful pen testing process, various solutions were presented using different techniques and tools. Different protective precautions and actions are already being taken into consideration in order to protect a given system from any cyber-attack based on ethical hacking techniques, rather than only training managers~\cite{hajdarevic2017training,Ethicalh92:online}. In~\cite{rushing2015collaborative}, Rushing et al. described a software project that surrounds the network pen testing alongside the problems with team-based efforts. It was achieved by using already available network analysis technologies and tools. Hence, the Collaborative Penetration-testing and Analysis Toolkit (CPAT) was introduced as a mean for security analysts to ensure a collaborative foundation to conduct pen testing especially to aid with the (data) reconnaissance phase. CPAT was presented as a solution to aid network security analysts to perform pen testing through the provision of collaborative environments along reactive data management.
Tetskyi et al.~\cite{tetskyi2018neural} presented their solution based on the use of Neural Network Based, as a choice of tools for Pen testing of Web Applications. Their web service design which uses a neural network in order to create the decision support tool. Despite its simple implementation and flexibility,  the main drawback of their solution is  the high number of requirements to suit the experts’ needs. Aside from this, the advantage is based on its simple implementation compared to other algorithms, alongside its flexibility. Shah et al.~\cite{shah2014automated} presented the “Net-Nirikshak 1.0” model developed to facilitate the VAPT~\cite{goel2015vulnerability} in Indian Banks. Such a  tool is fully automated and interactive and does not require any high technical skills nor expertise. Therefore, making it easy for Bank professionals to conduct the test themselves. This tool can report the severity level of each identified vulnerability, while also detecting and exploiting SQL injection vulnerabilities. A Comprehensive Online Tool [WR-3] that detects security flaws in networks was presented by Trivedi et al. in~\cite{trivedi2010comprehensive}. Such a  solution is a presented application that can be used in order to analyse any network security flaw. This allows network administrators to secure their networks in a convenient manner. The authors also stated that their provided IDS solution prevents any cross scripting vulnerability, whilst analysing any requested service(s) for any possible security flaw. PENTOS, or “Penetration Testing Tool for Internet of Thing Devices” was presented by the Visoottiviseth et al. in~\cite{visoottiviseth2017pentos}. PENTOS helped users in determining any vulnerability and risk that surround the IoT device. PENTOS was developed their system by combining several technologies including Kali Linux and Linux APIs for penetration testing. However, the authors stated that they haven’t finished their work based on ZigBee pen testing module due to its requirement for a special yet expensive specialised ZigBee sniffing hardware. In~\cite{zhou2012analysis}, Zhou et al. mentioned the ASCFETA, or” Analysis System for Computer Forensic Education, Training \& Awareness” system. ASCFETA is playing a very important and crucial role in the organisation’s security policy. The main goal of this system is to ensure the system’s analysis, which helps any organisation in establishing the most perfect ASCFETA plan. This offered the ability for developers and managers to know and identify any insufficiency in the computer forensic field. A new approach to ensure a better Vulnerability Assessment and Penetration Testing Accuracy, was presented by Goel et al. in~\cite{goel2016ensemble}. Such approach aims to increase the accuracy level of vulnerability assessment and pen testing. In fact, the authors explained VAPT techniques and tools, alongside their limitations, before implementing their own model whilst developing a software named “VEnsemble 1.0”. Such a software is operational on various number of VAPT tools, and can detect different vulnerability types through the combination of different types of open source VAPT tools. \\ 
Moreover, Ning et al.~\cite{ning2008design} presented a pen testing attack model that can effectively describe the relations between different attack types in a successive order of executed attack. This model also combines the situation of the target under attack, along with the attack execution. Results revealed that this model can effectively instruct the pen attack and unfiy its implementation. Therefore, it can be useful in intrusion detection domains. However, Sandhya et al. relied on the use of Wireshark as a packet sniffer technique to perform pen testing of information gathering in \cite{sandhya2017assessment}. Their use of this tool was based on indicating whether a website is secure or non-secure. The authors stated that results did show that an assessed website has a given lapse in its own security mechanism. On the other hand, a Threat Model Driven Approach for automated pen testing was presented by Almubairik et al. in~\cite{almubairik2016automated}. Such algorithm was successfully designed to evaluate the immunity of tested systems against malicious attacks. The algorithm was designed for human reading instead of machines using mathematical notations. The algorithm can also be enabled in business in order to take security countermeasures to minimise the impact of sensitive data exposures. Thus, ensuring business continuity by helping security experts avoid the overlooking for any threat. Thus, relieving testers from these hard tasks~\cite{angmo2014performance}. In~\cite{abbasi2014descriptive}, Abbasi et al. managed to identify expert hackers along their specialities, through a scalable and generalised framework that analyses the hacker’s forum content. Their framework is a social media analytical model that can be applied to various forms of User Generated Content (UGC) including structural and content features. Therefore, allowing the extraction of any interaction between users and hacking communities, including hackers' Web forum~\cite{lu2010social}. Therefore, contributing to social media analytics and cyber-security research alike. Gamification is a new technique that is being adopted by various fellow researchers to evaluate the effectiveness of a given security measure and evaluate its performance in a simulated attack scenario. 
\par
Unlike other solutions, an innovative approach based on an ethical game of hacking, protected by an authorisation infrastructure that observes user activity pre/post adaptation was presented by Bailey et al.~in~\cite{bailey2018evaluating}. Live experiments revealed the ability to handle malicious behaviour of real and intelligent users, while also capturing users response to new adaptation.

Moreover, Bechtsoudis et al. conducted a network pen testing using specific processes and tools that can scan the network for any possible vulnerability, and discover any exploitation mechanism in~\cite{bechtsoudis2012aiming}. These tests can either be conducted by internal IT security department, or by external certified penetration testing and security auditors. Their experiment ensured a successful man-in-the-middle attack through the Generic Routing Encapsulation (GRE) tunnel, where users’ interactions were caught using a sniffing machine. Results ensured that pen tests must be conducted regularly to the organisation’s network. J. Wang conducted a study over the nature of the mobile internet, using a network itself to build a given platform and complete the execution plan of mobile pen testing in~\cite{jiajia2016research}. Testing results revealed a high impact on the terminals of mobile communications, which accelerated the energy consumption. Moreover, if the system terminal and application vulnerability were combined, a high level of threats will be imposed. As a result, a pen testing model based on the mobile communication terminal was established, with experimental data being provided to ensure the extraction, detection and defence of features against any network attack. In~\cite{guzman2017iot}, a mobile pen testing specialised company named Attify, started giving courses, writing books and conducting live pen testing on different IoT aspects to ensure a much more secure and safe use of mobile devices.
\\

In their paper~\cite{alazab2013crime}, Alazab et al. managed to investigate the use of different crime toolkits including Zeus, Silent Banker, and SpyEye~\cite{sood2013dissecting} alongside other toolkits. It was also realised that malware can be created by non-technical attackers, which increases the number of malware types. This has increased the complexity in the malware’s quality. After proper investigation, the authors made a prediction over the rise of the complexity in overcoming these tools and means. More precisely, against the polymorphic variations of malwares, where signature-based anti-viruses are not adequate to it. Therefore, it became really essential to conduct a behavioural based analysis or white-listing as alternatives to increase the reliability and usefulness of these countermeasures. As a result, these toolkits will be presented in \tablename~\ref{tab:3}.
\begin{table*}[!ht]
\centering
\small
\caption{\textbf{Crimeware Toolkits}}
\label{tab:3}
\begin{tabular}{|l|l|}
\hline\\
\textbf{Crime Toolkits} & \textbf{Description} \\ \hline
Carberp  & stronger than Zeus, uses both general and targeted attacks \\ \hline
Cridex & embedded URL link that tricks the user into browsing on compromised websites \\ \hline
Haxdoor & Redirects the user’s infected URL connection requests \\ \hline
InfoStear & capable of collecting victim’s confidential information \\ \hline
Limbo & injects user’s webpage, and steals sensitive data \\ \hline
Silent Banker & occurs when users click on on a malicious link \\ \hline
SpyEye & observes and Sniffs HTTP, FTP and POP network protocols \\ \hline
URL Zone & Modifies bank pages to lure victims to make transfers on the fake webpages\\ \hline
Zeus & Uses traditional e-mail phishing methods\\ \hline
\end{tabular}
\end{table*}

\par
Aside these toolkits, these solutions can be summarized in \tablename~\ref{tab:31}.

\begin{table*}[!ht]
\centering
\small
\caption{{Pen Testing Existing Solutions}}
\label{tab:31}
\begin{tabular}{|p{1.5cm}|p{1.5cm}|p{2cm}|p{3cm}|p{8cm}|}
\hline
\textbf{Year} & \textbf{Reference} & \textbf{Authors} & \textbf{Solution} & \textbf{Details} \\ \hline
{2008} & {\cite{ning2008design}} & {Ning et al.} & {Pen Testing Attack Model} & {Effectively Describes Relations  Between  Different  Attack  Types } \\ \hline
{2010} & {\cite{trivedi2010comprehensive}} & {Trivedi et al.} & { Comprehensive  Online  Tool [WR-3]} & {Analyses Any Network Security Flaw} \\ \hline
{2012} & {\cite{zhou2012analysis}} & {Zhou  et  al.} & { ASCFETA } & {Part of An Organisation’s   Security   Policy} \\ \hline
{2013} & {\cite{alazab2013crime}} & {Alazab et al.} & {Crime Tool Kit Investigation} & {Study Of Malware Type, Variation \& Quality} \\ \hline
{2014} & {\cite{shah2014automated}} & {Shah et al.} & {“Net-Nirikshak 1.0”  model } & {Facilitates VAPT In Indian Banks} \\ \hline
{2014} & {\cite{abbasi2014descriptive}} & { Abbasi et al. } & {Social  Media  Analytical Model} & {Exctracts Interactions Between Users \& Hacking Communities} \\ \hline
{2015} & {\cite{rushing2015collaborative}} & {Rushing  et  al. } & {CPAT} & {Suitable For Pen Testing Data Reconnaissance Phase} \\ \hline
{2016} & {\cite{goel2016ensemble}} & {Goel et al.} & {“VEnsemble 1.0” Approach} & {A VAPT That Detects Different Vulnerability Types} \\ \hline
{2016} & {\cite{almubairik2016automated}} & {Almubairik et al.} & {Threat  Model  Driven  Approach} & {Evaluates The Immunity of Tested Systems Against Attack Types} \\ \hline
{2017} & {\cite{visoottiviseth2017pentos}} & {Visoottiviseth  et  al.} & {PENTOS} & {Determines Risks/Vulnerabilities Of IoT Devices} \\ \hline
{2017} & {\cite{sandhya2017assessment}} & {Sandhya et al.}& {Use of Wireshark} & {Assesses Website Security} \\ \hline
{2018} & {\cite{tetskyi2018neural}} & { Tetskyi  et  al. } & {Neural Network Based} & {Creates Decision Support Tool For Web Application Pen Testing} \\ \hline
{2018} & {\cite{bailey2018evaluating}} & { Bailey et al.} & {Ethical Game Of  Hacking} & {Handlee Malicious Behaviour Of Real \& Intelligent  Users} \\ \hline

\end{tabular}
\end{table*}


\section{Security \& Safety Procedures}
\label{sec:6}
Despite the different preventive and protective security measures being recommended, many steps need to be considered to ensure both security and safety measures. Incident responders must be distinguished and classified to maintain the right response against any given event(s). Once achieved, preventive and protective security measures must be employed for further protection. Finally, these security measures must be employed to maintain and ensure the right level of protection with respect to security goals.

\subsection{Security \& Safety Procedures Steps}
The focus should also be based on ensuring security and safety procedures that each employee must follow and adhere to. In fact, it can only be achieved by relying on continuous,  consecutive and constant training and awareness, including:
\begin{itemize}
\item \textbf{Step 1:} employees must not use Free Wi-Fi to log in, nor rely on public computers so their credentials are not stolen. Additionally, changing passwords regularly also helps. It is also important to know which site is being visited through the “secure sign” and “Hyper Text Transfer Protocol Secure” (HTTPS) instead of “HTTP”.
\item \textbf{Step 2:} the browser’s history must be cleared to prevent any attacker from conducting a cookie theft or/and masquerading attacks without the user’s knowledge. Moreover, it is also important to know what link to click on, simply by avoiding spam email based on phishing attack types.
\item \textbf{Step 3:}  IT staff must be trained as emergency response and disaster recovery teams in case of any possible attack taking place. Therefore, being able to assess the likelihood and impact of a given risk based on the threat of exploiting a given vulnerability.
\item \textbf{Step 4:} the company’s paper waste must be eliminated by destroying every piece of paper, disk, or hardware component beyond recovery and recognition, so that the thrown garbage cannot be used as an information gathering method to know the company’s little secrets. This can also be achieved if the organisation follows paperless processes.
\item \textbf{Step 5:} strong identification and authentication mechanisms can be used if they rely on biometric procedures to access sensitive locations in a given organisation, through additional extensive security procedures. Moreover, employees must lock their devices (i.e laptops, PCs, etc) after leaving for a break or home, and lock their desks to prevent any confidential paper leakage. Therefore, no device should be left unattended.
\item \textbf{Step 6:} employees must be trained against different social engineering, and phishing attack techniques and types. This can be done by limiting the information given on a given phone, or through face-to-face chatting, or even through instant messaging. This also includes how to identify phishing attacks based on sending fake infected CV formats (pdf, Microsoft word, etc..), or sending malicious links, or war-dialing and vishing attacks.
\item \textbf{Step 7:} employees must be trained in their own domains against various cyber-attack types, but mainly against insider attacks. Moreover, employees must be prevented from using USBs, in addition to their own devices. This step can prove itself as a useful method for that specific purpose.
\item \textbf{Step 8:} Previous employees should have all their forms of access controls, rights and privileges to the system removed. This includes their previous IDs, access privileges, passwords, e-mails, biometrics and cards.
\end{itemize}

\subsection{Incident Responders}
Usually responders try to respond back to the attackers. Such responses can either be active or passive. This depends on the responders’ skills, experience, available resources and manpower. As such, they can be classed as active, passive or hybrid.

\begin{itemize}
    
\item \textbf{Active Response:}
is characterised by the ability to “counter-back”any given attack against a system, organisation, government, or in some cases against certain individuals~\cite{himma2008ethical} . In fact, an active response can also trace back the source of the attack and neutralize it. It can also turn the occurring attack against the attacking source itself. The main attack types are known to be as DoS attacks and buffer overflow attacks. Such active response requires an advanced level of skills and knowledge along with  experience in the IT domain. This includes Information and Communication Technology (ICT) and Information Technology and Communication (ITC).

\item \textbf{Passive Response:}
 usually takes place when there are not enough skills to trace back a given attack or counter it back, but instead, it relies on taking defensive counter-measures to overcome it. Passive responses can take another aspect, where the attack source is being monitored over a long period of time before being targeted after gathering enough information about it.

\item \textbf{Hybrid Response:}
or "Smart Response", is based on implementing smart security measures that can reduce the burden on cyber-security experts. Such implementation is based on the employment of Machine Learning in Cyber-Security, IDS/IPS. Moreover, Honeypots and Honeynets are also being employed to track and prevent any unauthorised access or/and attack. 
\end{itemize}

\subsection{Preventive \& Protective Security Measures}
\begin{figure*}[!ht]
  \centering
    \includegraphics[scale=0.4]{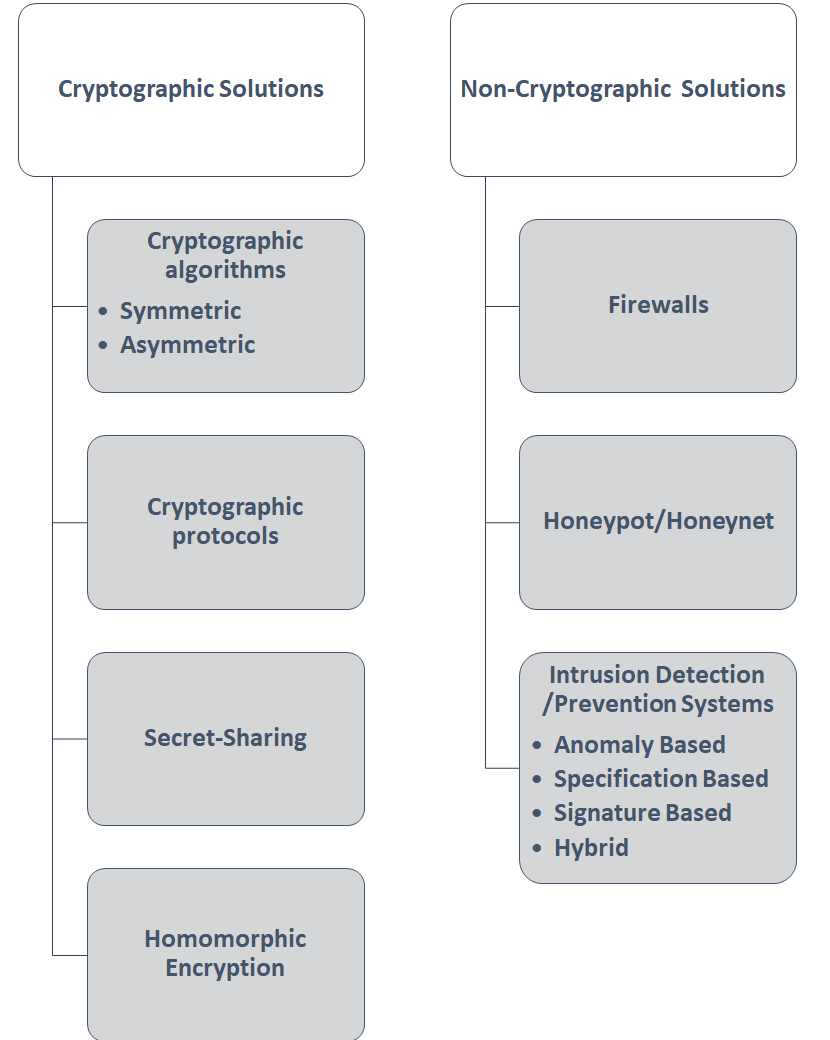}
    \caption{Traditional Security Measures}
    \label{fig:g}
\end{figure*}

As a precaution, different papers~\cite{wood2005implementing,brewer2015cyber} discussed the importance of implementing the necessary preventive and protective security measures to ensure a higher level of security of a given system. Hence, the proposed taxonomy is illustrated in \figurename~\ref{fig:g}.

\begin{itemize}
\item \textbf{Access Control:}
The need for access control mechanisms is a must. Depending on the employee’s role and the assigned task(s), the access privilege will be granted for a given amount and period of time, or until the task has been fulfilled and achieved. In fact, there are various access control mechanisms, except that they can be divided into three main classes as follows: 

\begin{itemize}
\item \textbf{Discretionary Access Control:} based on assigning a given employee the least privilege, in order to perform his intended task. It can be Role Based Access Control (RoBAC), Rule Based Access Control (RuBAC), Task Based Access Control (TBAC), and Attribute Based Access Control (ABAC).
\item \textbf{Mandatory Access Control:} or MAC. is based on assigning access control according to each user’s main role.
\item \textbf{Discretionary Access Control:} or DAC. is based on granting access to a given personnel, according to their post and status in a given organisation.
\end{itemize}

\item \textbf{Closing Open Ports:}
Closing open and unnecessary ports operating on the Transport Layer (User Datagram Protocol (UDP) and Transport Cntrol Protocol (TCP)), is recommended to reduce and tighten the security gap. This can prevent a given attacker from possibly exploiting it. 

\item \textbf{Firewalls:}
The need for smart firewalls (i.e Artificial-Intelligence (AI) based firewalls, Application Firewall and Next Generation Firewalls)is also recommended as a policy to prevent a private network from the untrusted internet. Therefore, allowing firewalls to monitor incoming and outgoing traffic packets and filter them by relying on stateful, stateless, dynamic or application firewalls, along with next generation, and zone policy firewall types. This either allows a given packet to be granted access, or denies its access with or without notifying the source.

\item \textbf{Anti-Virus:}
The need for effective anti-viruses and anti-malware is highly needed to detect any virus/malware. Advanced anti-viruses can detect advanced types of malware attacks. Therefore, deploying strong and capable auto-updatable anti-viruses allows an organisation to detect various malware types. In contrast, polymorphic malware and crypting services including zero-day attacks still remain as the main persistent challenge.

\item \textbf{Honeypots:}
are another smart technique and mechanism that can be employed. Based on sacrificing an unneeded system that serves as a trap, luring a given attacker with the impression of targeting a real device. Meanwhile, in the background, the attacker is trapped, and security personnel will be able to track and bring him down, as well as learning from his criminal behaviour. Honeynets can also be used. 

\item \textbf{Intrusion Detection/Preventing Systems:}
The employment of IDS/IPS systems relies on the organisation’s abilities to ensure either an active response, passive response or hybrid response according to its needs. Moreover, IDS can be centralised, distributed or even hybrid. IDSs can also be employed as network-based, host-based, or application-based. Furthermore, IDSs along with IPSs can be signature-based, anomaly-based, network-based, behaviour-based or even Hybrid-based. 

\item \textbf{Forensics:}
The need to have more knowledge about the (digital) forensic and anti-forensic domains is also highly recommended, especially in order to protect or/and recover logs and audit files from being altered, modified or/and deleted beyond recovery. This helps in protecting these logs, in addition to knowing how to retrieve deleted data especially if it was accidentally deleted, while also initiating an forensics investigation process.

\item \textbf{Machine Learning:}
The need for machine learning in cyber-security can also be a new aspect, capable of implementing Artificial Intelligence (AI) to learn about an attack, with an accurate reasoning about any future and possible attack(s). AIs can take off a huge burden from the IT staff and personnel, especially if they are supervised/semi-supervised. Machine Learning can also be employed in cryptography to maintain a higher level of encryption that offers a serious challenge for a given hacker to break it.

\item \textbf{Cryptography:}
encrypting open wireless communications would prevent any eavesdropping or/and interception attack(s), and would also ensure a secure storage of data and information alike. The choice of cryptographic techniques~\cite{schneier2007applied,kanade2009multi} rely on the value, volume, amount, importance, and size of the (real) data being dealt with and sorted. This choice also depends on the type of encryption being symmetric or asymmetric, as well as lightweight or heavyweight encryption. In fact, different encryption techniques can be used including differential privacy, secret sharing, data obfuscation, homomorphic encryption, secret sharing, hashing~\cite{bellare2004hash,krawczyk1997hmac}, symmetric/asymmetric, alongside other available techniques based on using Virtual Private Networks (VPN), Proxies and The Onion Router (TOR). In fact, quantum and post-quantum cryptography seems to be the future of cryptography~\cite{chen2016report,ding2019post,alagic2019status}.
\end{itemize}

\subsection{Maintaining Security Goals}
Pen testing is applied to perform, achieve, fulfil and maintain the necessary security goals. Such achievement allows a pen tester to evaluate the levels of security whilst also assessing the level of vulnerability based on the likelihood and impact of a given risk. In fact, it is tested against a cyber-attack that can either target the data/information of a given system, the system itself, or both. 

\begin{itemize}
\item \textbf{Confidentiality:}
Pen testing is performed to evaluate the confidentiality level, of which is fulfilled and achieved through testing the levels of encryption and preventing any possible information gathering attempt(s). Confidentiality is mainly targeted through social engineering, reverse engineering, and eavesdropping based on passive and active information gathering (replay attack), and the ability to stop it and overcome it.

\item \textbf{Integrity:}
 must also be tested in order to evaluate the level of security and immunity mainly against man-in-the-middle and meet-in-the-middle attacks, replay attacks, session hijacking, data diddling (unauthorised data alteration) and salami attacks (minor data security attacks). Therefore, pen testing is applied to ensure a higher level of integrity by mitigating any possible future attack  using detective and preventive security countermeasures.

\item \textbf{Availability:}
Many attacks target data or/and system’s availability. Such attacks can be DoS, DDoS attacks, jamming attacks, de-authentication attacks, replay attacks, electrical power attacks (especially in countries like Lebanon), and Server Room Environment attacks (based on humidity, heat, cold, fire or flood). These attacks can severely cripple and disrupt the ability of a system to maintain its availability. Hence, pen testing is being applied along with the necessary recommendation to maintain availability by isolating the server room both logically and physically, and relying on computational devices. Such devices can play a backup role in case the system went down, in order to maintain the availability of a given system in case of an emergency, while also maintaining a constant data backup to prevent its loss. Different security measures were already being applied to overcome DoS and DDoS attacks~\cite{senie1998network}.

\item \textbf{Authentication:}
in order to associate the level of authentication, pen testing is performed, since the authentication represents the first line of defence. Its purpose is to evaluate its security levels against password cracking attacks. This includes brute-force, dictionary, rainbow table~\cite{narayanan2005fast}, and birthday attacks, in addition to replay and bypass attacks. Therefore, it is recommended to constantly change a password from time to time, use a strong password policy, and rely on a multi-factor authentication process to ensure a better security level.

\item \textbf{Accuracy:} is another important aspect in the pen testing domain and scope. In fact, the accuracy of pen testing attack relies on its ability to conduct a real-case simulation attack (without causing any harm to the system) to show how ethical hackers can successfully perform their attack step-by-step. It also offers the ability of the implemented security measures (previously recommended by ethical hackers) to accurately detect and thwart any similar real event/attack.

\item \textbf{Privacy:} can be represented by both confidentiality and integrity. However, the main focus is on privacy itself. Such a task is based on the ability to perform simulation attacks led by an insider capable of proclaiming himself as a legitimate employee to inject a malicious package mainly through USBs (rubber-ducky). The aim is, either to retrieve confidential and sensitive credentials and information, or to gain a high privilege access (rootkits~\cite{chiang2007case}). In other cases, it can also include stealing confidential papers and information either from the organisation’s desks, or from their thrown garbage. This can have drastic effects against the organisation’s privacy. Hence, ethical hackers  need to cover every possible aspect/scenario and secure them both, before a malicious hacker successfully manages to exploit them. 

\item \textbf{Trust:} is more related to the relationship between an organisation and  ethical hackers. In fact, the organisation will have to fully trust the actions taken by ethical hackers in order to conduct their pen testing. However, it puts on more pressure on ethical hackers since all the eyes will be focused on them to perform and conduct their tasks. Moreover, trust is built by showing, explaining and revealing how a malicious hacker/attacker can break into the system, and perform a cyber-attack through exploitation. Therefore,  trust is built based on the cooperation and collaboration between ethical hackers and the organisation itself.\\
As a result, a second framework is presented to reveal how attacks impact  security goals, especially when the attacks are targeting the infrastructure, network, web services, or/and applications.
\end{itemize}
  \begin{figure*}[!ht]
  \centering
    \includegraphics[width=0.9\textwidth]{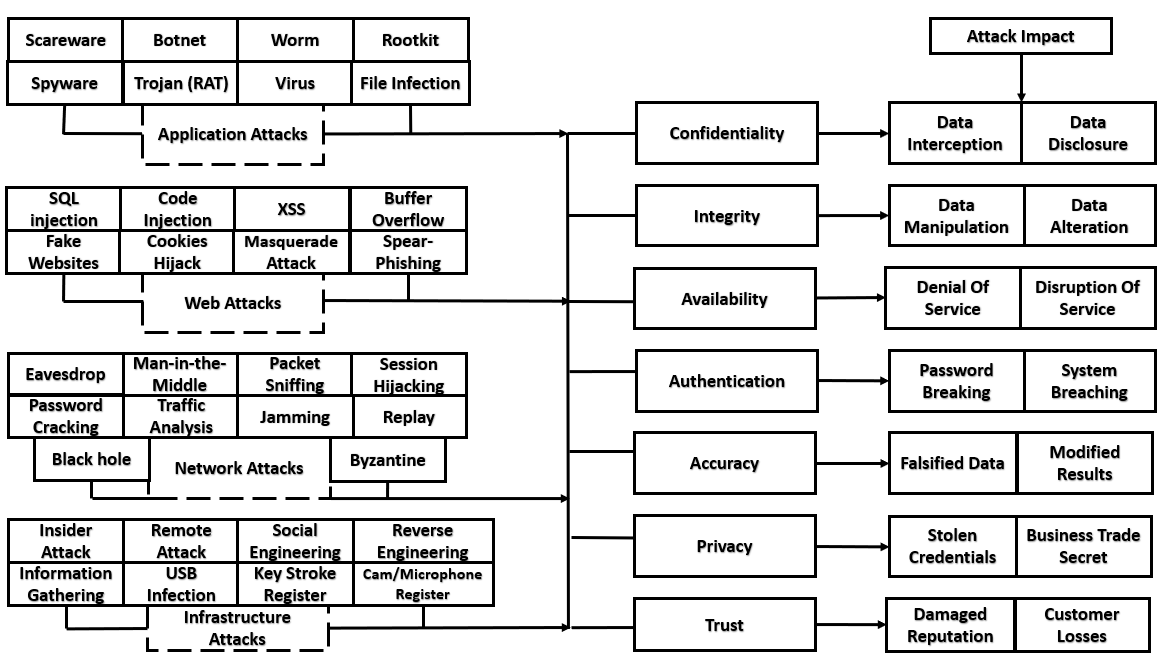}
    \caption{Framework Of Attacking Security Goals}
    \label{fig}
\end{figure*}

\section{Learnt Lessons \& Recommendations}
\label{sec:7}
After conducting this survey, the main weak points and strong points of pen testing are identified, along with those found in the ethical hacking domain. 
\begin{itemize}
\item \textbf{Higher Budget:} the need for a higher budget to train more ethical hackers is highly recommended to develop new skills and potentials, whilst also gaining new experience and knowledge. 
\item \textbf{Non-Uniform Training:} is also required in order to cover the field, personnel and security gaps as much as possible in the cyber-security domain. 
\item \textbf{Specialisation:} such a lesson would revoke the pressure off the ethical hackers by allowing them to focus and specialise in a specific field and domain.
\item \textbf{Forensics Knowledge:} the need for more knowledge about forensics and expertise is also required to conduct much more effective and efficient pen testing.
\item \textbf{Legal Authorisation:} laws must protect and support ethical hackers during their simulated attacks, where pre and post agreements are needed to be established first, to maintain their safety and security.
\end{itemize}

Despite having different security measures being available for implementation and deployment, this paper presents its own suggestions and recommendations to support ethical hacking and enhance pen testing domains.
\begin{itemize}
\item \textbf{More Specialised Tools:} are needed as part of vulnerability assessment and pen testing to ensure a higher accuracy in a real-time manner for heterogeneous systems/devices types. Therefore, a non-uniform training aspect is needed.

 \item \textbf{Fulfilling Skill-Gaps:} is caused due to the lack of available skilled ethical hackers. Hence, more advertisement and more recruitment campaigns, seminars and conferences are needed to encourage and educate more people to join the ethical hacking domain.
 
  \item \textbf{Specialised Training:} must be assigned for each staff/employee category to overcome and address to different attack types that they might encounter in their own fields.
  
  \item \textbf{Full Background Check \& Screening:} are required for each recruited staff before and during employment, especially those employed by a third party company responsible for cleaning the organisation's floors, desks, windows and different others, since these workers have various types of physical access privileges. 
 
  \item \textbf{Raising Awareness:} among employees to avoid any relevant document(s) being stolen or thrown in garbage, by destroying all sorts of papers through burning or using a paper-cutter (not very suitable).
 
  \item \textbf{More Workshops \& Resources:} are needed, especially books and electronic books and sources, as well as online courses and education to allow ethical hackers gain more knowledge and experience.
  
  \item \textbf{International Exchange \& Competitions:} are needed to allow more experience and skills to be gained through competence and gaming challenges, along the establishment of new fundamentals for (potential) ethical hackers to follow and adhere to.
  
\end{itemize}



\section{Conclusion}
\label{sec:9}
As a conclusion, the importance of conducting pen testing along with the reliance on trusted ethical hackers to perform simulated attacks are highlighted and discussed thoroughly. This allows the evaluation of the level of security against already known and common threats and attacks, alike. Therefore, ensuring a better understanding regarding the evaluation, assessment, and mitigation of a given risk. In fact, this survey is presented to give and offer a better insight over the difference between hacking and ethical hacking. Moreover, it also describes how hackers operate according to their objectives, goals, motives, gains, and benefits. Furthermore, various taxonomies and frameworks are presented as a background view and overview regarding the hacking domain. Finally, many recommendations and suggestions are proposed to further train employees, IT staff and security personnel, along with giving an insight about future work.
\par

A global point of view of pen testing was presented, along with the different and available techniques, technologies, approaches, and tools. As a result, the work will be based on a new, simple yet sophisticated ethical hacking tool that would ensure a high level of accuracy, whilst assessing risks and threats alike. This allows the proposed and presented security measures to adhere and overcome any possible malware attack, or threat. Therefore, the need for a semi-automated approach is a must and will be part of our future work.

\bibliographystyle{unsrt}

\end{document}